\documentclass[eqsecnum,floats, english,prd,aps,nofootinbib,superscriptaddress,amsfonts,longbibliography,tightenlines,preprint]{revtex4-1}
\usepackage[T1]{fontenc}
\usepackage[latin9]{inputenc}
\usepackage[english]{babel}
\usepackage{units}
\usepackage{amsmath}
\usepackage{esint}
\usepackage[unicode=true,pdfusetitle,
 bookmarks=true,bookmarksnumbered=false,bookmarksopen=false,
 breaklinks=false,pdfborder={0 0 1},backref=false,colorlinks=false]
 {hyperref}
\usepackage{breakurl}
\usepackage{color}
\usepackage{natbib}

\makeatletter
 
 \@ifundefined{textcolor}{}
 {
   \definecolor{BLACK}{gray}{0}
   \definecolor{WHITE}{gray}{1}
   \definecolor{RED}{rgb}{1,0,0}
   \definecolor{GREEN}{rgb}{0,1,0}
   \definecolor{BLUE}{rgb}{0,0,1}
   \definecolor{CYAN}{cmyk}{1,0,0,0}
   \definecolor{MAGENTA}{cmyk}{0,1,0,0}
   \definecolor{YELLOW}{cmyk}{0,0,1,0}
 }

\def\d{\textrm{d}}
\allowdisplaybreaks[1]

\makeatother

\begin{document}

\title{Constraint Lie algebra and local physical Hamiltonian\\
 for a generic 2D dilatonic model}

\author{Alejandro Corichi}

\email{corichi@matmor.unam.mx}

\affiliation{Centro de Ciencias Matem\'aticas, Universidad Nacional Aut\'onoma
de M\'exico, Campus Morelia, Apartado Postal 61-3, Morelia, Michoac\'an
58090, Mexico}

\affiliation{Center for Fundamental Theory, Institute for Gravitation and the Cosmos,
Pennsylvania State University, University Park
PA 16802, USA}

\author{Asieh Karami}

\email{karami@matmor.unam.mx}

\affiliation{Departamento de F\'{\i}sica, Universidad Aut\'{o}noma Metropolitana - Iztapalapa\\ San Rafael Atlixco 186, Mexico D.F. 09340, Mexico}
\affiliation{Centro de Ciencias Matem\'aticas, Universidad Nacional Aut\'onoma
de M\'exico, Campus Morelia, Apartado Postal 61-3, Morelia, Michoac\'an
58090, Mexico}

\author{Saeed Rastgoo}

\email{saeed@xanum.uam.mx}

\affiliation{Departamento de F\'{\i}sica, Universidad Aut\'{o}noma Metropolitana - Iztapalapa\\ San Rafael Atlixco 186, Mexico D.F. 09340, Mexico}
\affiliation{Centro de Ciencias Matem\'aticas, Universidad Nacional Aut\'onoma
de M\'exico, Campus Morelia, Apartado Postal 61-3, Morelia, Michoac\'an
58090, Mexico}

\author{Tatjana Vuka\v sinac}

\email{tatjana@umich.mx}

\affiliation{Facultad de Ingenier\'\i a Civil, Universidad Michoacana de San Nicol\'as
de Hidalgo, Morelia, Michoac\'an 58000, Mexico}

\date{\today}
\begin{abstract}
We consider a class of two dimensional dilatonic models, and revisit them from the perspective of a new set of ``polar type'' variables. These are motivated by recently defined variables within the spherically symmetric sector of 4D general relativity. We show that for a large class of dilatonic models, including the case {\it with} matter, one can perform a series of canonical transformations in such a way that the Poisson algebra of the constraints becomes a Lie algebra. Furthermore, we construct Dirac observables and a reduced Hamiltonian that accounts for the time evolution of the system. Thus, with our formulation, the systems under consideration are amenable to be quantized with loop quantization methods.

\end{abstract}
\maketitle

\section{Introduction}

The study of lower dimensional gravitational theories allows us to get a deeper insight into
various technically more involved problems of the four dimensional theory, such as black holes, some aspects of quantum gravity and 
the Hawking radiation. In two dimensions (2D) the Einstein-Hilbert action is a topological invariant and there are no 
local degrees of freedom, so additional fields are introduced in order to have a dynamical theory.  
The interest in general dilaton theories (GDT) in two dimensions was motivated principally by the
string inspired CGHS theory \cite{C.G.Callan1992} and the spherically symmetric reduction of 4D gravity \cite{Kuchar1994}. An extensive review of general 2D dilaton gravity theories, in the
first order formalism, can be found in \cite{Grumiller2002}. It has been shown that the classical GDT, without
matter fields, is exactly solvable, and all the classical solutions are found in this case \cite{Kloesch1996}. It turns out that
the theory is topological and there is a one parameter family of solutions labeled by a constant of motion. In
the presence of matter there are only few analytic solutions known \cite{Grumiller2002}. The careful analysis of asymptotic conditions allows the definition of the quasilocal energy for GDT as in \cite{Kummer_Lau1997}, for spherically symmetric gravity in \cite{Kuchar1994} and for the CGHS model in \cite{Varadaraj95}.
The canonical analysis and quantization of GDT was performed in
\cite{Louis-Martinez1994, Louis-Martinez1997} (see also \cite{Kuchar1997} for the CGHS model coupled to a scalar field).
The Hawking radiation in CGHS theory has been investigated in details, see, for example, \cite{C.G.Callan1992, A.Ashtekar2011}.
A recent result for the Hawking radiation of a spherical loop quantum gravity black hole has been presented in 
\cite{Gambini_Pullin2014}.

Generic dilaton theories in 2D are usually formulated in terms of a metric, a dilaton field and some additional matter or gauge
fields. They can be recast into the first order formalism, namely one can consider diads and $SO(1,1)$ connection as basic
gravitational variables, and include theories with non vanishing torsion \cite{Grumiller2002}. Following  ideas motivated by 
loop quantum gravity one can further introduce Ashtekar type variables, as a generalization of the variables obtained
in spherically symmetric reduction of 4D gravity \cite{BSvars05,us-new-var}. It turns out to be useful and convenient
to re-write GDT in these new variables in order to explore the possible loop quantization techniques. Recently, there have been
some new results based on this approach, such as the loop quantization of the Schwarzschild black hole \cite{Gamb_Pull2013},
where the corresponding quantum spacetime has been constructed and then used to analyze the Hawking radiation 
\cite{Gambini_Pullin2014}. One of the technical results that allowed the completion of the Dirac quantization procedure within the
LQG approach was the Abelianization of the algebra of the Hamiltonian constraints. This was achieved in the spherically
symmetric model \cite{Gamb_Pull2013}, as well as in CGHS case \cite{Rastgoo2013}, but it has not been explicitly performed in the generic case. 
One of the purposes of this work is to show that this can indeed achieved for general dilaton theories,
by performing a globally well defined scaling of the Lagrange multipliers. We also construct a true Hamiltonian that
governs the dynamics in a reduced phase space, in a generic case. Following the ideas put forward in \cite{Rovelli1991,Giesel2010,Thiemann2007} 
we construct Dirac observables and the physical Hamiltonian that governs their evolution, by 
interpreting a scalar matter field as a physical clock. 

The first step in this analysis is the selection of the basic variables in the Hamiltonian 
formulation of GDT. In order to motivate
the choice for canonical variables for a generic 2D dilaton gravity model that we use in this work (as introduced in
\cite{us-new-var}), let us recall the form of the 
Ashtekar type variables, for spherically symmetric spacetimes.
As we mentioned above, the Hamiltonian formulation of 4D gravity can be performed in terms
of metric or tetrad variables. In the former case, the fundamental degrees
of freedom of the gravitation field, in the canonical approach, are the induced 3-metric and its 
conjugate, $(q_{ab},p^{ab})$. 

The first order formalism, can be reformulated in terms of several
different pairs of canonical variables. One of them are the Ashtekar variables, that were introduced in
an attempt to obtain  constraints that are polynomial and first class, and the theory
can be formulated as a gauge theory so that one can apply Yang-Mills methods to it. The basic \emph{real}  Ashtekar variables
are the $su(2)$ valued Ashtekar-Barbero connection 
$A_{a}^{i}=\frac{1}{2}\epsilon^{i}{}_{jk}\omega_{a}{}^{jk}+\gamma K_{a}^{b}e_{b}^{i}$
 and its conjugate momentum, the densitized triad
$E_{i}^{a}=ee_{i}^{a}=\frac{1}{2}\epsilon_{ijk}\epsilon^{abc}e_{b}^{j}e_{c}^{k}$.
Here $e_{b}^{j}$ are the components of the triad field, $\omega_a^{jk}$ are the components of the spin connection 
compatible with the triad, $K_{a}^{b}$ 
are the components the extrinsic curvature of a space-like leaf of a foliation,
 $e$ is the triad determinant and $\gamma$ is the real-valued Barbero-Immirzi
parameter. Indices $\{i,j,k\}$ are the $SU(2)$ indices while $\{a,b,c\}$
are the spatial ones.

In the case where the system has spherical symmetry, the connection
can be expanded in the one-form basis $\{\d x,\d\theta,\d\phi\}$ and
the triad in the vector basis $\{\partial_{x},\partial_{\theta},\partial_{\phi}\}$.
The components of $A$ or $E$ in the $x$ direction,  $A^{x}$ and $E^{x}$, are 
a scalar density and 
a scalar, respectively. 
The angular components of $A$ are written as a combination of two
scalars $A_{1}$ and $A_{2}$, while the angular components of $E$
are the combination of the scalar densities $E_{1}$ and $E_{2}$.
Since the components $1$ and $2$ of the Gauss constraint are now
identically vanishing, the gauge group of the theory is now
reduced from $SU(2)$ to $U(1)$. One can introduce
a new set of variables that are invariant under $U(1)$ gauge symmetry, 
$A_{\varphi}^{2}=  A_{1}^{2}+A_{2}^{2}$ and 
$\left(E^{\varphi}\right)^{2}=  E_{1}^{2}+E_{2}^{2}$.
It turns out that $E^{\varphi}$ is not canonically conjugate to $A_{\varphi}$,
but that $(K_{\varphi},E^{\varphi})$ do form a canonical pair, where $K_{\varphi}$ is
the $\varphi$ component of the extrinsic curvature one-form. The new set of variables
$\{K_{x},E^{x},K_{\varphi},E^{\varphi}\}$ were introduced by Bojowald-Swiderski, and shall be called in what follows \emph{polar-type} variables \cite{BSvars05}.


We shall show that one can generalize the polar-type
variables to the case of generic 2D dilaton model.
It is worth noting that as we will see later, the physical interpretation
of these variables  might
be different in each submodel (for example CGHS vs spherically symmetric)
and this is something that one should be careful about especially
in quantization, but the computational methods can nevertheless be
extended to the whole system.


The Hamiltonian analysis of the general 2D dilaton model with matter, in the first order formalism, has been performed in \cite{Grumiller_PhD}. There the author showed that
the original first class constraint algebra can be redefined and abelianized, in the case without matter fields.  One of the new constraints turned out to be proportional to a spatial derivative of the ADM mass.  The Hamiltonian analysis 
of the generic 2D dilatonic gravity model
in metric variables, in the case without matter and without the kinetic term for the dilaton field, has been presented  in \cite{Louis-Martinez1994}. 
It has been shown that for an appropriate parametrization of the metric and dilaton variables, the Hamiltonian constraint that generates the evolution along the Killing vector field can be written as a spatial derivative of the phase-space 
function that represents the energy of the system. 
This property has also been noticed in some particular models. For example,  
in a different approach, new canonical variables for spherically
symmetric vacuum gravity were introduced, one of them being the mass as a function of the radius \cite{Kuchar1994}. 
The gravitational part of the Hamiltonian constraint in that case is given by the total derivative of the mass
with respect to the radial variable.  It has been shown that the set of original first class constraints,
that satisfy the hypersurface deformation algebra is equivalent to a pair of simpler constraints  and one of them is given by the derivative of the mass with respect to the radial variable.
The corresponding result has been obtained in \cite{Varadaraj95} for the CGHS model. 
We shall show here that an analogue result can be obtained in generalized polar-type variables, for the general theory {\it with} matter, that 
leads to the Abelianization of the algebra  of the smeared Hamiltonian constraints. As far as we know, this is the first time that this result has been generalized to dilaton models with matter.  

The structure of this paper is the following. In Sec.~\ref{Sec1} we introduce
the general 2D dilatonic model coupled to a scalar field, and perform the Hamiltonian analysis in terms of diads and a $SO(1,1)$ connection, the so called Cartan variables. 
We explicitly treat two possible cases regarding the
presence of kinetic term for a dilaton field, since it enables us to be more flexible and to include more models.
In Sec.~\ref{Sec2} we introduce a procedure to derive the generalized polar-type variables for a whole
generic class of 2D dilatonic systems and rewrite the constraints in terms of these new variables. In Sec.~\ref{Sec3} we
provide a prescription to Abelianize the Hamiltonian constraint. In Sec.~\ref{Sec4} we construct a set of Dirac observables
and show that a scalar field can be used as a physical time variable suitable for describing the evolution of
this constrained system. We also construct a physical Hamiltonian that governs this evolution. In Sec.~\ref{Sec5} we
present conclusions and some indications for future work.

\section{Preliminaries}
\label{Sec1}

\subsection{General 2D dilaton theories}

Quantizing  full general relativity has proven to be a quite difficult challenge and
there are still some unsolved issues in various approaches to this problem.
On the other hand there is the general expectation that, until finding a full theory of
quantum gravity, we should still be able to learn some important
aspects of such a theory
by studying simpler models,
such as symmetry reduced ones or lower dimensional toy models. It
turns out that there is a generic class of 2D dilatonic theories that
contains some of these important symmetry reduced and toy models.
Moreover, formulating this generic system classically in such a way that it will
be suitable for quantization is an important task, 
and this is the main purpose of this work.

To start, we shall remind the reader that the most general diffeomorphism invariant action yielding second order
differential equations for the metric $g_{ab}$ and a scalar (dilaton)
field $\Phi$ in two dimensions coupled to a scalar matter field $f$,
is \citep{Grumiller2002} 
\begin{equation}
S=\int \d^{2}x\sqrt{-g}\left(Y(\Phi)R(g)+V\left(\left(\nabla\Phi\right)^{2},\Phi\right)\right)-\int \d^{2}x\sqrt{-g}\, W(\Phi)g^{ab}\partial_{a}f\partial_{b}f\, ,\label{eq:most-generic-action}
\end{equation}
where $g=\det(g_{ab})$ and $Y(\Phi),\, V\left(\left(\nabla\Phi\right)^{2},\Phi\right)$ and $W(\Phi)$ are model dependent functions
of the dilaton field \citep{Rastgoo2013}. There have been extensive studies of such systems and there is a long list of references 
that can be found, for instance, in \citep{Grumiller2002}.

In this class we choose a subclass \citep{Banks1991,Odintsov1991,Kloesch1996}
that is general enough for our purposes,
\begin{equation}
S=\int \d^{2}x\sqrt{-g}\left(Y(\Phi)R(g)+\frac{1}{2}g^{ab}\partial_{a}\Phi\partial_{b}\Phi+V(\Phi)\right)-\int \d^{2}x\sqrt{-g}\, W(\Phi)g^{ab}\partial_{a}f\partial_{b}f.\label{eq:Generic-Action}
\end{equation}

In the above action, the kinetic term $\frac{1}{2}g^{ab}\partial_{a}\Phi\partial_{b}\Phi$
can be removed 
by performing a conformal transformation
\begin{equation}
\tilde{g}_{ab}=\Omega^{2}(\Phi)g_{ab},\label{eq:conformal-trans}
\end{equation}
with
\begin{equation}
\Omega(\Phi)=C\exp\left(\frac{1}{4}\int \d\Phi\frac{1}{\frac{\d Y(\Phi)}{\d\Phi}}\right)\, ,\label{eq:Omega-integ}
\end{equation}
and $C$ being a suitable constant of integration \citep{us-new-var}.
We have introduced the possibility of this elimination because some
of the specific models that we are interested in can only be derived
from the above action if we remove this term. An example of this is
the 3+1 spherically symmetric model in polar-type
variables \cite{BSvars05}, 
although one should note that in this case, $\Phi$ is not a dilaton field but it is 
the $g_{\theta\theta}$ component of the metric  
in the spherically symmetric ansatz.
Some other models such as the well-known CGHS model can be derived from
(\ref{eq:Generic-Action}) with the kinetic term present \citep{us-new-var,Rastgoo2013}. For a brief list of some 
important theories that can be written using (\ref{eq:most-generic-action}) see Appendix \ref{app:models-table}.

\subsection{The generic $2D$ Hamiltonian in Cartan variables}

To find the Hamiltonian in the generalized polar-type  variables \citep{BSvars05}, we first
need to follow the usual process in LQG, namely first write the theory
in the Cartan variables (diads and spin connection), ADM decompose it and then make a Legendre transformation
to find the Hamiltonian. From there we shall make a canonical transformation
to generalized polar-type variables, guided by what has been done for the
3+1 spherically symmetric and the CGHS model in \citep{us-new-var,Rastgoo2013}.

We shall start by writing the metric in diads 
\begin{equation}
g_{ab}=\eta_{IJ}e^{I}{}_{a}e^{J}{}_{b}\, ,
\end{equation}
where $\eta_{IJ}$ is the Minkowski metric, $e^{I}{}_{a}$ are the diads
and $I,J=\{0,1\}$ are the internal Lorentz indices while $a,b$ are
the abstract (spacetime) ones. In 2D the spin connection only has two components 
$\omega_{a}{}^{IJ}=\omega_{a}\epsilon^{IJ}$, and the curvature tensor only has one independent component,
and we will take it to be the scalar curvature $R=2\partial_{[a}\omega_{b]}\epsilon^{IJ}e^{a}{}_{I}e^{b}{}_{J}$.
General 2D first order gravity theories can have a non vanishing torsion \cite{Grumiller2002}, but we are interested
in the torsionless case, so we should impose it as an additional condition,
\begin{equation}
T^I_{ab}:=\left(\d e^{I}+\epsilon^{I}{}_{J}\omega\wedge e^{J}\right)_{ab}=2\partial_{[a}e_{b]}{}^{I}+2\epsilon^{I}{}_{J}\omega_{[a}e_{b]}{}^{J}=0\, .
\label{torsionless}
\end{equation}
Now, the gravitational-dilaton part of the Lagrangian density (\ref{eq:Generic-Action}) in Cartan variables takes the following form
\begin{align}
L_{\textnormal{g}} &= -X_I\epsilon^{ab}T^I_{ab}+e\bigl( Y(\Phi )R
\underbrace{+ \frac{1}{2}\eta_{IJ}e^{I}{}_{a}e^{J}{}_{b}\partial_{a}\Phi\partial_{b}\Phi}_{\textrm{kin}} + V(\Phi )\bigr)\nonumber \\
&=-2X_{I}\epsilon^{ab}(\partial_{[a}e_{b]}{}^{I}+\epsilon^{I}{}_{J}\omega_{[a}e_{b]}{}^{J})+2Y\partial_{[a}\omega_{b]}e\epsilon^{IJ}e^{a}{}_{I}e^{b}{}_{J}\underbrace{+\frac{1}{2}\eta^{IJ}ee_{I}{}^{a}e_{J}{}^{b}\partial_{a}\Phi\partial_{b}\Phi}_{\textrm{kin}}+eV.\label{eq:act-tet1}
\end{align}
where,  $e=\det(e_{a}{}^{I})$, $\epsilon^{ab}=-ee_I{}^a e_J{}^b\epsilon^{IJ}$ and
$\underbrace{\ldots}_{\textrm{kin}}$ refers to the terms that are only present if the kinetic term in (\ref{eq:Generic-Action})
is present. We have introduced new fields $X_I$ in order to impose the vanishing of the torsion condition. It is easy to see that the variation
of the action with respect to $X_I$ leads to (\ref{torsionless}). 
Integrating by parts (and assuming that the boundary term vanishes\footnote{In this work we shall not specify boundary 
conditions, but we should bear in mind that a careful treatment of boundary conditions is necessary in order to have a well
posed action (differentiable and finite), and to define an energy and conserved charges of the theory. See, for example, 
\cite{Kuchar1994, Varadaraj95} for definition of energy in spherically symmetric and CGHS models, respectively.}) 
in the first term in Lagrangian density (\ref{eq:act-tet1})
will give us the pure gravitational Lagrangian density as 
\begin{align}
L_{\textrm{g}}= & e\left(-2\partial_{a}(X_{I})e_{K}{}^{a}\epsilon^{KI}-2X_{I}e^{Ia}\omega_{a}+2Y\partial_{[a}\omega_{b]}\epsilon^{IJ}e_{I}{}^{a}e_{J}{}^{b}\underbrace{+\frac{1}{2}\eta^{IJ}e_{I}{}^{a}e_{J}{}^{b}\partial_{a}\Phi\partial_{b}\Phi}_{\textrm{kin}}+V\right).\label{eq:act-tet-ger}
\end{align}
The matter part of the Lagrangian density in (\ref{eq:Generic-Action}), can
also simply be written as 
\begin{equation}
L_{\textnormal{m}}=-W\eta^{IJ}ee_{I}{}^{a}e_{J}{}^{b}\partial_{a}f\partial_{b}f.\label{eq:act-tet-m}
\end{equation}
Next, we decompose the Lagrangian density by ADM method and perform a Legendre
transformation to get to the Hamiltonian density. Most of the details needed
for these steps have been already discussed in  \citep{us-new-var},
so we just recall the results here. We assume that the spacetime is foliated by surfaces $\Sigma_t$, parametrized by $t$. 
The configuration variables are $(^{*}X^{I},\omega_1,\Phi ,f)$, where $^{*}X^{I}=\epsilon^{IJ}X_{J}$ is the
Hodge dual of $X^{I}$ and their corresponding canonical momenta are $(P_I,P_{\omega},P_{\Phi}, P_f)$, with
\begin{align}
P_{I}= & \frac{\partial L}{\partial{}^{*}\dot{X}^{I}}=2\sqrt{q}n_{I},\label{eq:PI1}\\
P_{\omega}= & \frac{\partial L}{\partial\dot{\omega}_{1}}=2Y(\Phi),\label{eq:Pomega1}\\
P_{f}= & \frac{\partial L}{\partial\dot{f}}=-\frac{2W(\Phi)\sqrt{q}}{N}\left(N^{1}f'-\dot{f}\right),\label{eq:Pf1}
\end{align} 
and
\begin{align}
P_{\Phi}=& \frac{\partial L}{\partial\dot{\Phi}}= 0,\ \ \ \ \ \ \ \ \ \ \ \ \ \ \ \ \ \ \ \ \ \textrm{if the kinetic term is absent}\label{eq:Pphi2}\\
P_{\Phi}= & \frac{\partial L}{\partial\dot{\Phi}}=\frac{\sqrt{q}}{N}\left(N^{1}\Phi'-\dot{\Phi}\right),\,\,\,
\textrm{if there is the kinetic term}\label{eq:Pphi1}
\end{align}  
Here $N$ is lapse, $N^{1}$ is the (one dimensional) shift vector, 
$n_{I}=\eta_{IJ}{e^J}_a n^a$, where $n^a$ is the unit timelike normal to
the spatial hypersurface $\Sigma_t$, $q_{ab}$ is the induced spatial metric on $\Sigma_t$, $q$ is its determinant, 
and the prime represents partial derivative with respect to the spatial coordinate $x^{1}=x$.

In the case when the kinetic term is absent there is a pair of second class primary constraint $\mu_1:= P_{\omega}-2Y(\Phi)$ and
$\mu_2:=P_{\Phi}$, so we should pass to the corresponding Dirac brackets and afterwards treat these constraints as identities on
the phase space. As a result, $\Phi =Y^{-1}(\frac{P_{\omega}}{2})$ and $P_{\Phi} =0$ and the Dirac brackets of the remaining phase
space variables reduce to the Poisson brackets. In the case where the kinetic term is present, we can express $\dot{\Phi}$ from
the equation (\ref{eq:Pphi1}), as a function of canonical variables, and the pair $(\Phi , P_{\Phi})$ is not eliminated from
the phase space of the system. We still have the primary constraint $\mu_1$. 

We note that (\ref{eq:PI1}) is also a primary constraint but since the momentum conjugate to $n_{I}$ is zero, i.e. 
$P_{n_I}=\partial L/\partial\dot{n}_{I}=0$, and this is also a constraint that makes a second class pair with 
(\ref{eq:PI1}), they can be solved together to yield $n_{I}=P_{I}/2\sqrt{q}$ and, as a consequence
\begin{equation}
\Vert P\Vert=\sqrt{-|P|^{2}}=\sqrt{-\eta^{IJ}P_{I}P_{J}}=\sqrt{-4q\eta^{IJ}n_{I}n_{J}}=2\sqrt{q},\label{eq:P-vert-norm}
\end{equation}
where  $\Vert P\Vert$ is the norm of $P_{I}$.

The corresponding Hamiltonian density then becomes\footnote{In this work the Hamiltonian and diffeomorphism constraints are denoted
by $\mathcal{H}_{0}$ and $\mathcal{H}_{1}$ respectively, while $\mathcal{H}_{0}(N)=N\mathcal{H}_{0}$
and $\mathcal{H}_{1}(N^{1})=N^{1}\mathcal{H}_{1}$ will be used for
their corresponding densities. The smeared version of each density will
be written as $H_{0}(N)=\int dx\,N\mathcal{H}_{0}$ and $H_{1}(N^{1})=\int dx\,N^{1}\mathcal{H}_{1}$
and finally the total Hamiltonian density is denoted by $H=\mathcal{H}_{0}(N)+\mathcal{H}_{1}(N^{1})+c^{i}\Upsilon_{i}$
where $\Upsilon_{i}$ are other constraints if there are any, and $c^{i}$
are Lagrange multipliers.}
\begin{align}
H= & N\left[2\frac{P_{1}}{\Vert P\Vert}\left(^{*}X^{0}\right)'+2\frac{P_{0}}{\Vert P\Vert}\left(^{*}X^{1}\right)'-2\frac{P_{0}}{\Vert P\Vert}\omega_{1}{}^{*}X^{0}-2\frac{P_{1}}{\Vert P\Vert}\omega_{1}{}^{*}X^{1}\right.\nonumber \\
 & \underbrace{-\frac{1}{\Vert P\Vert}\Phi^{\prime2}-\frac{P_{\Phi}^{2}}{\Vert P\Vert}}_{\textrm{kin}}+\frac{2W(\Phi)f^{\prime2}}{\Vert P\Vert}+\frac{P_{f}^{2}}{2W\Vert P\Vert}-\frac{\Vert P\Vert}{2}V(\Phi)\bigg]\nonumber \\
 & +N^{1}\left[\underbrace{P_{\Phi}\Phi'}_{\textrm{kin}}+P_{f}f'+P_{0}\left(^{*}X^{0}\right)'+P_{1}\left(^{*}X^{1}\right)'-P_{0}{}^{*}X^{1}\omega_{1}-P_{1}{}^{*}X^{0}\omega_{1}\right]\nonumber \\
 & +\omega_{0}\left[P_{0}{}^{*}X^{1}+P_{1}{}^{*}X^{0}-\left(2Y(\Phi)\right)^{\prime}\right]+B\underbrace{\left(P_{\omega}-2Y(\Phi)\right)}_{\textrm{kin}}.\label{eq:H3}
\end{align}
Note that $N,N^1,B$ and $\omega_0$ are Lagrange multipliers but $X^I$, which in the beginning entered the theory as a Lagrange multiplier, is now promoted to a canonical variables due to the integration by parts that we performed in (\ref{eq:act-tet-ger}).  


The Hamiltonian density (\ref{eq:H3}) is a linear combination of 
constraints. These are
the scalar or Hamiltonian constraint multiplied by $N$, the vector
or diffeomorphism constraint multiplied by $N^{1}$ and the Gauss
constraint multiplied by $\omega_{0}$. As we mentioned, if the kinetic
term is present, we will have another constraint, $\mu_1:= P_{\omega}-2Y(\Phi)$. 
This case will be considered in section \ref{sub:with-kinetic-H}.

\section{The Hamiltonian in generalized polar-type variables}
\label{Sec2}

In this section we show how the generalized polar-type variables can be derived for the 
theories given by (\ref{eq:H3}). We shall treat separately the cases with and without a kinetic term,
since the number and nature of the canonical variables and the constraints are different in these two cases.  
We shall perform a series of canonical transformations in order to obtain a simplified form for the constraints in the new variables.

\subsection{The case without the kinetic term}

In this case the Hamiltonian density is (\ref{eq:H3}) without the terms corresponding
to the presence of the kinetic term. This means that we have a conformal
transformation (\ref{eq:conformal-trans}) and thus a conformal factor
$\Omega(\Phi)$ in our formulation such that
\begin{equation}
\tilde{q}=\tilde{q}_{11}=\tilde{g}_{11}=\Omega^{2}(\Phi)g_{11},\label{eq:q-Omega-g11}
\end{equation}
since our spatial metric is one dimensional. Also note that in this
case $\Phi$ is not a dynamical variable since its time derivative does not appear in the Lagrangian. 

From (\ref{eq:P-vert-norm}) and (\ref{eq:q-Omega-g11}) we get
\begin{equation}
\Vert P\Vert=2\sqrt{\tilde{q}}=2\Omega(\Phi)\sqrt{g_{11}}.\label{eq:P-Vert-Omega}
\end{equation}
Now since our model has only one spatial dimension, it has only
one independent spatial metric component. Let us call that
\begin{equation}
E_{1}:=\sqrt{g_{11}}.\label{eq:E1-g11}
\end{equation}
Let us also introduce a new variable $E_2$ as
\begin{equation}
E_2:=2P_{\omega}=4Y(\Phi )\, ,\label{eq:E2-Y}
\end{equation}
where in the last line we used the primary second class constraint (\ref{eq:Pomega1}).
From this relation we can express $\Phi =Y^{-1}(\frac{E_2}{4})$, only when $Y(\Phi )$ has an inverse
function. In various models $Y(\Phi )=k\Phi^2$ (where $k$ is a constant), as can be seen in appendix \ref{app:models-table}.
So in this case, in order to have an inverse function we should restrict to the region $\Phi\ge 0$ or $\Phi\le 0$.        
Now from (\ref{eq:P-Vert-Omega})-(\ref{eq:E2-Y}) we can write
\begin{equation}
\Vert P\Vert= 
2E_{1}\Omega(E_{2}),\label{eq:can-trans-PVert}
\end{equation}
and using this, we can define a canonical transformation to new momenta 
$E_{1}$, $E_{2}$ and $\eta$ as
\begin{align}
P_{0}= & 2E_{1}\Omega(E_{2})\cosh{\eta},\label{eq:can-trans-P0}\\
P_{1}= & 2E_{1}\Omega(E_{2})\sinh{\eta},\label{eq:can-trans-P1}\\
P_{\omega}= & \frac{E_{2}}{2},\label{eq:can-trans-POm}
\end{align}
where $\eta$ is a gauge angle. 

We have a canonical transformation from the original set of canonical variables
in the gravitational sector $(q,p):=\{( ^{*}X^{I},\omega_1),(P_I,P_{\omega})\}$ to a new set of
the canonical variables $(Q,P):=\{(K_1,A_2,Q_{\eta}),(E_1,E_2,\eta )\}$ and
it can be seen from (\ref{eq:can-trans-PVert})-(\ref{eq:can-trans-POm})
that the corresponding generating function is 
\begin{equation}
G(q,P)=2{}^{*}X^{0}E_{1}\Omega(E_{2})\cosh{\eta}+2{}^{*}X^{1}E_{1}\Omega(E_{2})\sinh{\eta}+\omega_{1}\frac{E_{2}}{2}.
\end{equation}
Then from this, the new canonical variables $(K_{1},A_{2},Q_{\eta})$ conjugate to the momenta $(E_{1},E_{2},\eta )$
are
\begin{align}
Q_{\eta}= & \frac{\partial G}{\partial\eta}=2E_{1}\Omega(E_{2})\left(^{*}X^{0}\sinh{\eta}+{}^{*}X^{1}\cosh{\eta}\right)={}^{*}X^{0}P_{1}+{}^{*}X^{1}P_{0},\\
K_{1}= & \frac{\partial G}{\partial E_{1}}=2\Omega(E_{2})\left(^{*}X^{0}\cosh{\eta}+2{}^{*}X^{1}\sinh{\eta}\right)=\frac{^{*}X^{0}P_{0}+{}^{*}X^{1}P_{1}}{E_{1}},\\
A_{2}= & \frac{\partial G}{\partial E_{2}}=\frac{2E_{1}\Omega^{\prime}(E_{2})}{E_{2}^{\prime}}\left(^{*}X^{0}\cosh{\eta}+{}^{*}X^{1}\sinh{\eta}\right)+\frac{\omega_{1}}{2}=\frac{\Omega^{\prime}(E_{2})}{E_{2}^{\prime}\Omega(E_{2})}\left({}^{*}X^{0}P_{0}+{}^{*}X^{1}P_{1}\right)+\frac{\omega_{1}}{2}.
\end{align}
Using these and the definition of the new momenta (\ref{eq:can-trans-PVert})-(\ref{eq:can-trans-POm}), the original variables $^{*}X^{0}, ^{*}X^{1},\omega_1$ in terms of the new canonical variables are 
\begin{align}
\omega_{1}= & 2\left(A_{2}-\frac{E_{1}K_{1}\Omega^{\prime}(E_{2})}{E_{2}^{\prime}\Omega(E_{2})}\right),\\
^{*}X^{0}= & \frac{K_{1}\cosh{\eta}}{2\Omega(E_{2})}-\frac{Q_{\eta}\sinh{\eta}}{2E_{1}\Omega(E_{2})},\\
^{*}X^{1}= & \frac{Q_{\eta}\cosh{\eta}}{2E_{1}\Omega(E_{2})}-\frac{K_{1}\sinh{\eta}}{2\Omega(E_{2})}.
\end{align}
Substituting these into the Hamiltonian density (\ref{eq:H3}) yields
\begin{align}
H= & N\left[\frac{Q_{\eta}'}{E_{1}\Omega(E_{2})}-\frac{E_{1}^{\prime}Q_{\eta}}{E_{1}^{2}\Omega(E_{2})}-\frac{\Omega^{\prime}(E_{2})Q_{\eta}}{E_{1}\Omega^{2}(E_{2})}-\frac{2K_{1}}{\Omega(E_{2})}\left(A_{2}+\frac{1}{2}\eta^{\prime}\right)+\frac{2E_{1}K_{1}^{2}\Omega^{\prime}(E_{2})}{E_{2}^{\prime}\Omega^{2}(E_{2})}\right.\nonumber \\
 & +\left.\frac{W(E_{2})\left(f^{\prime}\right)^{2}}{E_{1}\Omega(E_{2})}+\frac{P_{f}^{2}}{4W(E_{1})E_{1}\Omega(E_{2})}-E_{1}\Omega(E_{2})V(E_{2})\right]\nonumber \\
 & +N^{1}\left[-2Q_{\eta}\left(A_{2}+\frac{1}{2}\eta^{\prime}\right)+K_{1}^{\prime}E_{1}+P_{f}f'
 +\frac{K_{1}E_{1}\Omega^{\prime}(E_{2})}{\Omega(E_{2})}+\frac{2K_{1}E_{1}\Omega^{\prime}(E_{2})Q_{\eta}^{\prime}}{E_{2}^{\prime}\Omega(E_{2})}\right]\nonumber \\
 & +\omega_{0}\left[Q_{\eta}-\frac{1}{2}E_{2}^{\prime}\right].
\end{align}
We can define a new variable
\begin{equation}
K_{2}=A_{2}+\frac{1}{2}\eta^{\prime},\label{def_K2}
\end{equation}
and substitute it in the above Hamiltonian density. 
In the following we shall fix the gauge symmetry
by introducing the gauge fixing condition
\begin{equation}
\eta=1.\label{eq:eta-fix}
\end{equation}
It turns out that this condition is second class together with the Gauss constraint.
Therefore, we can solve the Gauss constraint to get
\begin{equation}
Q_{\eta}=\frac{1}{2}E_{2}^{\prime}.\label{eq:gauss-solve}
\end{equation}
Then, from (\ref{eq:eta-fix}) it follows that $K_2=A_2$. Substituting (\ref{eq:eta-fix}) and (\ref{eq:gauss-solve})
back into the Hamiltonian density yields
\begin{align}
H= & N\left[\frac{E_{2}^{\prime\prime}}{2E_{1}\Omega(E_{2})}-\frac{E_{1}^{\prime}E_{2}^{\prime}}{2E_{1}^{2}\Omega(E_{2})}-\frac{\Omega^{\prime}(E_{2})E_{2}^{\prime}}{2E_{1}\Omega^{2}(E_{2})}+\frac{2E_{1}K_{1}^{2}\Omega^{\prime}(E_{2})}{E_{2}^{\prime}\Omega^{2}(E_{2})}\right.\nonumber \\
 & \left.-\frac{2K_{1}K_{2}}{\Omega(E_{2})}+\frac{W(E_{2})\left(f^{\prime}\right)^{2}}{E_{1}\Omega(E_{2})}+\frac{P_{f}^{2}}{4W(E_{2})E_{1}\Omega(E_{2})}-E_{1}\Omega(E_{2})V(E_{2})\right]\nonumber \\
 & +N^{1}\left[-E_{2}^{\prime}K_{2}+K_{1}^{\prime}E_{1}+f^{\prime}P_{f}\right]
 := N\mathcal{H}_0 + N^1 \mathcal{H}_1 .\label{eq:H-noKin-BojoVar}
\end{align}
The canonical pairs 
\begin{equation}
\left\{ \left(K_{1},E_{1}\right),\left(K_{2},E_{2}\right)\right\} 
\end{equation}
are analogues of the polar-type variables $\{ (K_x,E^x),(K_{\varphi},E^{\varphi})\}$ for this
generic system, and therefore we will refer to them as the {\it generalized  polar-type variables}. 
For example, comparing the above Hamiltonian density with the one for the 3+1 spherically symmetric case 
in \citep{us-new-var}, we see that the polar-type variables for
the 3+1 spherically symmetric case correspond to the the following
change of variables (canonical transformation)
\begin{align}
K_1=&\sqrt{E^{x}} K_{\varphi}\, ,\ \ \ E_{1}=  \frac{E^{\varphi}}{\sqrt{E^{x}}},\\ \nonumber
K_2 =& K_x\, ,\ \ \ \ \ \ \ \ \ E_{2}=  E^{x},
\end{align}
with $\Omega(E_{2})=  \left(E^{x}\right)^{\nicefrac{1}{4}}$.
The densitized triads $E^x$ and $E^\varphi$ are the radial and angular components of the 
momentum conjugate to the connection \citep{BSvars05}. Thus, the Hamiltonian density 
(\ref{eq:H-noKin-BojoVar}) is the Hamiltonian density 
of a generic 2D dilatonic system without the kinetic term written
in generalized polar-type variables.
Let us now consider the case when there is a kinetic term.

\subsection{The case with the kinetic term\label{sub:with-kinetic-H}}

Since there is no conformal transformation in this case, we do not
have a conformal factor $\Omega$ to consider. Furthermore, there is a new primary constraint
that, from (\ref{eq:Pomega1}), reads
\begin{equation}
\mu_{1}=P_{\omega}-2Y(\Phi)\approx0.
\end{equation}
Thus for this case we have the following Hamiltonian density which is the
sum of (\ref{eq:H3}) with terms associated to the presence of kinetic
term, and the above new primary constraint
\begin{align}
H= & N\left[2\frac{P_{1}}{\Vert P\Vert}\left(^{*}X^{0}\right)'+2\frac{P_{0}}{\Vert P\Vert}\left(^{*}X^{1}\right)'-2\frac{P_{0}}{\Vert P\Vert}\omega_{1}{}^{*}X^{0}-2\frac{P_{1}}{\Vert P\Vert}\omega_{1}{}^{*}X^{1}\right.\nonumber \\
 & -\frac{1}{\Vert P\Vert}\Phi^{\prime2}-\frac{P_{\Phi}^{2}}{\Vert P\Vert}+\frac{2W(\Phi)f^{\prime2}}{\Vert P\Vert}+\frac{P_{f}^{2}}{2W\Vert P\Vert}-\frac{\Vert P\Vert}{2}V(\Phi)\bigg]\nonumber \\
 & +N^{1}\left[P_{\Phi}\Phi'+P_{f}f'+P_{0}\left(^{*}X^{0}\right)'+P_{1}\left(^{*}X^{1}\right)'-P_{0}{}^{*}X^{1}\omega_{1}-P_{1}{}^{*}X^{0}\omega_{1}\right]\nonumber \\
 & +\omega_{0}\left[P_{0}{}^{*}X^{1}+P_{1}{}^{*}X^{0}-\left(2Y(\Phi)\right)^{\prime}\right]+B\left[P_{\omega}-2Y(\Phi)\right],\label{eq:H4}
\end{align}
where $B$ is a new Lagrange multiplier. 
Following the ideas of the last subsection, let us name $E_{1}$ the variable corresponding
to the only independent spatial metric component
\begin{equation}
E_1:=\sqrt{g_{11}}=\sqrt{q_{11}}=\sqrt{q}\, .\label{eq:E1-q}
\end{equation}
Furthermore, we define $E_2$ as
\begin{equation}
E_2:= P_{\omega}\, .
\end{equation}

Thus, from (\ref{eq:P-vert-norm}) and (\ref{eq:E1-q}) we obtain
\begin{equation}
\Vert P\Vert= 2\sqrt{q}=2E_{1},
\end{equation}
so we can make the following change of variables to the new momenta $(E_{1}, E_{2}, \eta )$
\begin{align}
P_{0}= & 2E_{1}\cosh{\eta},\\
P_{1}= & 2E_{1}\sinh{\eta},\\
P_{\omega}= & E_{2}.\label{eq:PO-E2}
\end{align}
Like in the previous case we can find a generating function for this canonical transformation
\begin{equation}
G(q,P)=2{}^{*}X^{0}E_{1}\cosh{\eta}+2{}^{*}X^{1}E_{1}\sinh{\eta}+\omega_{1}E_{2},
\end{equation}
yielding the following results for the canonical variables $(K_{1},A_{2}, Q_{\eta})$ 
conjugate to the momenta $(E_{1}, E_{2}, \eta )$ as
\begin{align}
Q_{\eta}= & \frac{\partial G}{\partial\eta}=2E_{1}\left(^{*}X^{0}\sinh{\eta}+{}^{*}X^{1}\cosh{\eta}\right)={}^{*}X^{0}P_{1}+{}^{*}X^{1}P_{0},\\
K_{1}= & \frac{\partial G}{\partial E_{1}}=2\left(^{*}X^{0}\cosh{\eta}+2{}^{*}X^{1}\sinh{\eta}\right)=\frac{^{*}X^{0}P_{0}+{}^{*}X^{1}P_{1}}{E_{1}},\\
A_{2}= & \frac{\partial G}{\partial E_{2}}=\omega_{1}.
\end{align}
Using the above relations, one can write the old canonical variables
in terms of the new ones as
\begin{align}
\omega_{1}= & A_{2},\\
^{*}X^{0}= & \frac{K_{1}\cosh{\eta}}{2}-\frac{Q_{\eta}\sinh{\eta}}{2E_{1}},\\
^{*}X^{1}= & \frac{Q_{\eta}\cosh{\eta}}{2E_{1}}-\frac{K_{1}\sinh{\eta}}{2}.
\end{align}
Now if we substitute the new variables in the Hamiltonian density (\ref{eq:H4})
and define the new variable $K_2$ as in (\ref{def_K2}),
we get
\begin{align}
H= & N\left[\frac{Q_{\eta}^{\prime}}{E_{1}}-\frac{Q_{\eta}E_{1}^{\prime}}{E_{1}^{2}}-K_{1}K_{2}-\frac{\Phi^{\prime2}}{2E_{1}}-\frac{P_{\Phi}^{2}}{2E_{1}}+\frac{W(\Phi)(f^{\prime})^{2}}{E_{1}}+\frac{P_{f}^{2}}{4W(\Phi)E_{1}}-E_{1}V(\Phi)\right]\nonumber \\
 & +N^{1}\left[E_{1}K_{1}^{\prime}-Q_{\eta}K_{2}+\Phi^{\prime}P_{\Phi}+f^{\prime}P_{f}\right]+\omega_{0}\left[Q_{\eta}-\left(2Y(\Phi)\right)^{\prime}\right]
 +B\left[E_{2}-2Y(\Phi)\right].
\end{align}

We can again find a gauge fixing condition that fixes the gauge angle $\eta=1$ as in  (\ref{eq:eta-fix}) and, considering that this is second class 
together with the Gauss constraint, solve the Gauss constraint as
\begin{equation}
Q_{\eta}=2Y^{\prime}(\Phi).
\end{equation}
Substituting these in the Hamiltonian density yields
\begin{align}
H= & N\left[\frac{2Y^{\prime\prime}(\Phi)}{E_{1}}-\frac{2Y^{\prime}(\Phi)E_{1}^{\prime}}{E_{1}^{2}}-K_{1}K_{2}-\frac{\Phi^{\prime2}}{2E_{1}}-\frac{P_{\Phi}^{2}}{2E_{1}}+\frac{W(\Phi)(f^{\prime})^{2}}{E_{1}}+\frac{P_{f}^{2}}{4W(\Phi)E_{1}}-E_{1}V(\Phi)\right]\nonumber \\
 & +N^{1}\left[E_{1}K_{1}^{\prime}-2Y^{\prime}(\Phi)K_{2}+\Phi^{\prime}P_{\Phi}+f^{\prime}P_{f}\right]+B\left[E_{2}-2Y(\Phi)\right].\label{eq:H5}
\end{align}
Again, we have arrived at the canonical pairs 
\begin{equation}
\left\{ \left(K_{1},E_{1}\right),\left(K_{2},E_{2}\right)\right\} 
\end{equation}
which are the generalized polar-type variables for the generic case with
a kinetic term present.

\subsubsection{Second class procedure}

In this case, since we have an additional primary constraint which in the new variables is expressed as
\begin{equation}
\mu_{1}=E_{2}-2Y(\Phi)\approx0,\label{eq:mu1}
\end{equation}
one should check its consistency, i.e. that it is preserved under evolution
\begin{equation}
\dot{\mu}_{1}=\left\{ \mu_{1},\int \d  xH\right\} \approx0.
\end{equation}
Doing so reveals that we have another new (secondary) constraint
\begin{equation}
\dot{\mu}_{1}\approx0\Rightarrow\mu_{2}=K_{1}+\frac{2P_{\Phi}}{E_{1}}\frac{d Y(\Phi)}{d\Phi}\approx0.
\end{equation}
The preservation of $\mu_{2}$ does not lead to any new constraints. It turns out that these two constraints 
are second class together and thus, for this case, we should follow the Dirac procedure for the second
class systems. This means that we first should solve these two constraints,
and then consider  the Dirac brackets instead of the Poisson brackets.

Solving these two constraints yields
\begin{equation}
\mu_{1}=0\Rightarrow E_{2}=2Y(\Phi),\label{eq:mu1-sol}
\end{equation}
For the second one, first note that from above and using the formula
for the derivative of the inverse functions we have
\begin{equation}
2\frac{\d}{\d E_{2}}Y^{-1}\left(\frac{E_{2}}{2}\right)=\frac{1}{\frac{\d}{\d\Phi}Y\left(\Phi\right)},
\end{equation}
so solving $\mu_{2}=0$ yields
\begin{equation}
\mu_{2}=0\Rightarrow P_{\Phi}=-K_{1}E_{1}Z(E_2),\label{eq:mu2-sol}
\end{equation}
where we have used the notation
\begin{equation}
Z(E_2)=\frac{\d}{\d E_{2}}Y^{-1}\left(\frac{E_{2}}{2}\right).
\end{equation}

Substituting $Y(\Phi)$ and $P_{\Phi}$ from (\ref{eq:mu1-sol}) and
(\ref{eq:mu2-sol}) into (\ref{eq:H5}) yields
\begin{align}
H= & N\left[\frac{E_{2}^{\prime\prime}}{E_{1}}-\frac{E_{1}^{\prime}E_{2}^{\prime}}{E_{1}^{2}}-K_{1}K_{2}-\frac{Z^2(E_{2})E_{2}^{\prime2}}{2E_{1}}-\frac{K_{1}^{2}E_{1}Z^2(E_{2})}{2}+\frac{W(E_{2})(f^{\prime})^{2}}{E_{1}}+\frac{P_{f}^{2}}{4W(E_{2})E_{1}}-E_{1}V(E_{2})\right]\nonumber \\
 & +N^{1}\left[E_{1}K_{1}^{\prime}-E_{2}^{\prime}K_{2}-K_{1}E_{1}E_{2}^{\prime}Z^2(E_{2})+f^{\prime}P_{f}\right].
 \label{eq:H-tot-2nd-preD}
\end{align}
The next step in the second class procedure is to introduce the Dirac
brackets and use them instead of the Poisson ones. In this case, the general form
of the Dirac brackets for any two phase space function $A$ and $B$ is
\begin{equation}
\{A(x),B(y)\}_{D}=\{A(x),B(y)\}-\int \d w\int \d z\left(\{A(x),\mu_{i}(w)\}C^{ij}(w,z)\{\mu_{j}(z),B(y)\}\right),
\end{equation}
where $C^{ij}$ 's are the elements of the inverse matrix of $C_{ij}(x,y)=\{\mu_{i}(x),\mu_{j}(y)\}$
where in this case are
\begin{equation}
C_{ij}^{-1}(x,y)=C^{ij}(x,y)=\begin{pmatrix}0 & E_{1}Z^2(E_{2})\\
-E_{1}Z^2(E_{2}) & 0
\end{pmatrix}\delta(x-y).
\end{equation}
Using this, the Dirac brackets among canonical variables and momenta take the following form
\begin{align}
\{K_{1}(x),E_{1}(y)\}_{D} & =\{f(x),P_{f}(y)\}_{D}=\delta(x-y),\label{eq:Dirac-Br-pairs}\\
\{K_{2}(x),K_{1}(y)\}_{D} & ={\displaystyle K_{1}Z^2(E_{2})\delta(x-y)},\label{eq:Dirac-Br-KxKf}\\
\{K_{2}(x),E_{2}(y)\}_{D} & =-E_{1}Z^2(E_{2})\delta(x-y),\label{eq:Dirac-Br-KxEf}
\end{align}
\begin{equation}
\{E_{2},K_{1}\}_{D} =\{E_{1},E_{2}\}_{D}=\{f,\textrm{any except }P_{f}\}_{D}=\{P_{f},\textrm{any except }f\}_{D}=0.
\label{eq:Dirac-Br-other}
\end{equation}
It can thus be seen that the Dirac brackets are not in the 
``standard canonical form''. 
In the next section we shall use a simple prescription to bring the Dirac brackets to the
standard from.

\section{Constraint Lie algebra}
\label{Sec3}

As is well known, the Hamiltonian and diffeomorphism constraints satisfy the 1+1 dimensional surface deformation
algebra, which is not a Lie algebra. This feature might in turn lead to difficulties when attempting the Dirac quantization
procedure. However, it turns out that by an appropriate redefinition of the constraints, 
the algebra of Hamiltonian constraints can be made 
Abelian in the generic case, and as a result the whole algebra becomes a Lie algebra. This result
has already been proven in special cases, such as spherically symmetric vacuum gravity \cite{Gamb_Pull2013},
in polar-type coordinates, and it was an important step in the loop quantization of this model.
Here, we shall show how to achieve the Abelianization of the Hamiltonian constraint in the general 2D dilaton 
theory, by explicitly performing the rescaling of the lapse and shift functions in this general case.

\subsection{The case without the kinetic term}
\label{Sec3.1}

Here, as a first step, we rescale the shift 
\begin{equation}
\bar{N}^{1}=N^{1}+N\frac{2K_{1}}{E_{2}'\Omega(E_{2})}.\label{scale_shift1}
\end{equation}
After this rescaling $K_2$ disappears from the Hamiltonian constraint
in (\ref{eq:H-noKin-BojoVar}). 
It turns out that if we also rescale the lapse as 
\begin{equation}
\bar{N}=N\frac{E_{1}\Omega(E_{2})}{E_{2}^{\prime}},\label{scale_lapse1}
\end{equation}
the total Hamiltonian density (\ref{eq:H-noKin-BojoVar}) will become
\begin{equation}
H = \bar{N}\bar{\mathcal{H}}_0 + \bar{N}^{1}\mathcal{H}_1\, ,
\label{total_hamiltonian_without_kinetic_term}
\end{equation}
where $\bar{N}$ and $\bar{N}^{1}$ are new arbitrary (phase-space independent) lapse and shift functions. 
The new Hamiltonian constraint is given by
\begin{equation}
\bar{\mathcal{H}}_0 = \partial_{x} {\mathcal C}[E_1,K_1,E_2]
 +\frac{1}{E_{1}^{2}\Omega^{2}(E_{2})}
 \biggl(W(E_{2})E_{2}^{\prime}f^{\prime2} +\frac{E_{2}^{\prime}P_{f}^{2}}{4W(E_{2})}-2f^{\prime}P_{f}K_{1}E_1\biggr)\, ,
\label{h0_without_kinetic_term}
\end{equation}
where ${\mathcal C}$ is a phase space function in the gravitational-dilaton sector, defined as
\begin{equation}
{\mathcal C} [E_1,K_1,E_2]:=  \bigg(\frac{E_{2}^{\prime2}}{4E_{1}^{2}\Omega^{2}(E_{2})}-\frac{K_{1}^{2}}{\Omega^{2}(E_{2})}-
\int V(E_{2})\textrm{d}E_{2}\bigg)\, .\label{C}
\end{equation}
In the case without matter, ${\mathcal C}$ is a constant on constraint surface and commutes with all the first class constraints, hence being a global observable. In CGHS for example, it is the ADM energy of the system (see appendix \ref{app:comparison} and \cite{Gegenberg1995}).
The diffeomorphism constraint preserves its form,
\begin{equation}
\mathcal{H}_1  = -E_{2}^{\prime}K_{2}+K_{1}^{\prime}E_{1}+f^{\prime}P_{f}\, .
\label{h1_without_kinetic_term}
\end{equation}

Thus, the pure gravitational-dilaton part of the total Hamiltonian constraint 
becomes a total derivative, $\mathcal{H}_0^g=\partial_x{\mathcal C}$, an analogue of the result obtained in 
\cite{Louis-Martinez1994} for metric variables. 
As a consequence, the algebra of two smeared gravitational Hamiltonian constraints, 
$\bar{H}_{0}^{g}(N)=\int dx N(x)\bar{\mathcal{H}}_0^g(x)$, becomes Abelian,
\begin{equation}
\left\{\bar{H}_{0}^{g}(N),\bar{H}_{0}^{g}(M)\right\}=0\, ,
\label{eq:vac-Abel}
\end{equation}
as shown in the appendix \ref{sub:tot-deriv+matter-append}. Note that, as usual, we use the same notation for the smearing functions
$N$ and $N^1$, as for the lapse and shift functions, though generally they belong to different functional spaces. 

It can be explicitly shown that the algebra of smeared Hamiltonian constraints becomes Abelian also in the presence of matter 
(the details are presented in 
the  appendix \ref{sub:tot-deriv+matter-append})
\begin{equation}
\left\{\bar{H}_{0}(N),\bar{H}_{0}(M)\right\}=0\, .\label{eq:matt-Abel}
\end{equation}

With this, the algebra of the smeared Hamiltonian constraint $\bar{H}_{0}(N)$ and the smeared diffeomorphism constraint 
$H_{1}(N^1)=\int dxN^{1}(x)\mathcal{H}_{1}(x)$, becomes a Lie algebra
\begin{equation}
\begin{split}\left\{ H_{1}(N^{1}),H_{1}(M^{1})\right\} = & \ H_{1}\left(N^{1}(x)\partial_{x}M^{1}(x)-M^{1}(x)\partial_{x}N^{1}(x)\right)\, ,\\
\left\{ H_{1}(N^{1}),\bar{H}_{0}(N)\right\} = & \ \bar{H}_{0}\left(N^{1}(x)\partial_{x}N(x)\right)\, ,\\
\left\{ \bar{H}_{0}(N),\bar{H}_{0}(M)\right\} = & \ 0\, .
\end{split}\label{eq:Lie-alg-symb}
\end{equation}

Before we pass to the case with the kinetic term, let us make some comments about $\mathcal{C}$. It is known that the solutions of a generic 2D dilatonic model (\ref{eq:Generic-Action}), without the kinetic term and without matter have at least one Killing vector of the form \cite{Gegenberg1995, Grumiller_McNees}
\begin{equation}
k^a =\epsilon^{ab}\frac{\d Y}{\d \Phi}\, \partial_b \Phi 
= \frac{1}{4}\epsilon^{ab} \partial_b E_2\, ,
\end{equation}
where we have used the definition (\ref{eq:E2-Y}), $E_2=4Y(\Phi )$.
Then, from (\ref{eq:H-noKin-BojoVar}) it follows that $\dot{E}_2=N\frac{2K_1}{\Omega(E_2)} + N^1E_2'$, so 
the norm of the Killing vector is easily calculated
\begin{equation}
|k|^2 = \frac{1}{16\,\Omega^2(E_2)E_1^2}\, 
(E_2'^2-4E_1^2 K_1^2)\, .
\end{equation}
Now we can re-write the phase space function $\mathcal{C}$ as
\begin{equation}
{\mathcal C} [E_1,K_1,E_2]:=  \bigg(4|k|^2 -
\int V(E_{2})\textrm{d}E_{2}\bigg)\, .\label{C}
\end{equation}
This is the analogue of the result obtained in \cite{Gegenberg1995}
in different set of canonical variables.
Note that the hypersurfaces $E_2'=\pm 2 E_1 K_1$ in the phase space along which the norm of the Killing vector vanishes correspond to the Killing horizon. 

One might ask whether the rescalings of the lapse and shift that we performed are defined globally in phase space, or if there are restrictions on them that make them defined only locally.
We see from (\ref{scale_shift1}) and (\ref{scale_lapse1}), that the rescaling is well defined in the region of the phase space where $E_{2}^{\prime}\ne 0$, so we restrict the initial data for $E_2$ to
a space of monotonous functions of $x$. It follows that in the case without matter, when there is a Killing vector field, the rescaling is well defined even on the Killing horizon.   

\subsection{The case with the kinetic term}

In this case, the first step is to bring the Dirac brackets to the
standard form. By a simple inspection of (\ref{eq:Dirac-Br-KxKf}) and (\ref{eq:Dirac-Br-KxEf}),
one can see that by introducing  a new variables $U_{2}$
instead of $K_{2}$ as
\begin{equation}
U_{2}=K_{2}+E_{1}K_{1}Z^2(E_2),
\end{equation}
one can immediately obtain the Dirac brackets in canonical from as
\begin{align}
\{K_{1}(x),E_{1}(y)\}_{D}= & \delta(x-y)\, ,\\ \nonumber
\{U_{2}(x),E_{2}(y)\}_{D}= & \delta(x-y)\, ,\\ \nonumber
\{f(x),P_{f}(y)\}_{D}= & \delta(x-y)\, ,
\end{align}
while the remaining Dirac brackets vanish. 
This way the total Hamiltonian density  (\ref{eq:H-tot-2nd-preD})
becomes
\begin{align}
H= & N\bigg(\frac{E_{2}^{\prime\prime}}{E_{1}}-\frac{E_{1}^{\prime}E_{2}^{\prime}}{E_{1}^{2}}-K_{1}U_{2}-\frac{Z^2(E_{2})E_{2}^{\prime2}}{2E_{1}}+\frac{K_{1}^{2}E_{1}Z^2(E_{2})}{2}\\ \nonumber
 & +\frac{W(E_{2})(f^{\prime})^{2}}{E_{1}}+\frac{P_{f}^{2}}{4W(E_{2})E_{1}}-E_{1}V(E_{2})\bigg)\\ \nonumber
 & +N^{1}\left(E_{1}K_{1}^{\prime}-E_{2}^{\prime}U_{2}+f^{\prime}P_{f}\right).\label{H-kin-BojoVar-final}
\end{align}
Now, again, we can eliminate $U_{2}$ (like $K_{2}$ in the previous case)
from the Hamiltonian constraint 
by a rescaling of the shift as
\begin{equation}
\bar{N}^{1}=N^{1}+N\frac{K_{1}}{E_{2}^{\prime}}.\label{scale_shift2}
\end{equation}
The next step regarding the rescaling of the lapse function is a slightly more involved trickier than the previous case. 
By looking at the form of the Hamiltonian
constraint after the above rescaling, and using a bit of `educated guessing' regarding
what should be the suitable form of the terms in order to get total
derivatives, we rescale the lapse as
\begin{equation}
\bar{N}=N\frac{E_{1}}{A(E_{2})E_{2}^{\prime}},\label{scale_lapse2}
\end{equation}
where $A(E_{2})$ is an integration factor that allows us to express 
the pure gravitational part of the Hamiltonian constraint as a spatial derivative.
In order to find $A(E_{2})$ one separates the constraint
into the terms with $K_{1}$ and the ones without it, and demands that both sets of terms
be spatial derivatives, then both conditions lead to differential equations
yielding the same solution for $A$, given by
\begin{equation}
A(E_{2})=C_{0}\exp\left\{ -\int \d E_{2}\,Z^2(E_2)\right\} ,\label{A}
\end{equation}
with $C_{0}$ being a constant of integration. In this way we get for
the total Hamiltonian density
\begin{align}
H= & \bar{N}\left(\partial_{x}\left[\frac{E_{2}^{\prime2}A(E_{2})}{2E_1^2}-\frac{K_{1}^{2}A(E_{2})}{2}-\int A(E_{2})V(E_{2})\d E_{2}\right]\right.\nonumber \\
 & \left.+A(E_{2})\left[
 W(E_{2})E_{2}^{\prime}f^{\prime2}+\frac{P_{f}^{2}E_{2}^{\prime}}{4W(E_{2})}-
 \frac{f^{'}P_{f}K_{1}}{E_{1}}\right]\right)\nonumber \\
 & +\bar{N}^{1}\left(E_{1}K_{1}^{\prime}-U_{2}E_{2}^{\prime}+f^{'}P_{f}\right),
\label{Total_Hamiltonian_with_kinetic_term}
\end{align}
which is now written in the desired form. Once again, since the vacuum Hamiltonian constraint is now written as a spatial derivative, 
it strongly Poisson commutes with itself  both in the 
vacuum and in the cases with matter, as in (\ref{eq:vac-Abel}) and (\ref{eq:matt-Abel}).
Therefore, the algebra of constraints is a Lie algebra like in (\ref{eq:Lie-alg-symb}).

%

Also in this case, when there is no matter, there is a Killing vector that now is of the form \cite{Grumiller_McNees}
\begin{equation}
k^a=e^{-\frac{1}{2}E_2}\, \epsilon^{ab}\partial_b E_2\, .
\end{equation}
Its norm is given by
\begin{equation}
|k|^2=\frac{1}{4E_1^2}\, e^{-\frac{1}{2}E_2}(E_2'^2-E_1^2 K_1^2)\, .
\end{equation}


From (\ref{scale_shift2}) and (\ref{scale_lapse2}) we see that the rescaling is well defined when
$E_{2}^{\prime}\ne 0$, as well as $A(E_{2})\ne 0$. The second condition is fulfilled, as can be seen from
(\ref{A}), whenever $Z(E_2)$ is well defined (that is, whenever the inverse $Y^{-1}$ exists.) Note that, in the case without matter, again there is no obstruction to rescalings of the Lagrange multipliers on the Killing horizon. 

\section{Local physical Hamiltonian}
\label{Sec4}

In this section we shall construct the physical Hamiltonian that governs the evolution in the reduced phase space, in both cases, 
with or without the kinetic term. First step is the identification of the reduced phase space. We shall consider two approaches, 
the first one is gauge fixing and construction of the reduced Hamiltonian, and the second one, as in \cite{Giesel2010}, is the 
construction of Dirac observables and the physical Hamiltonian. 
The choice of the physical degrees of freedom or the corresponding Dirac observables is not unique, and we will make it based on the form of the constraints.
We shall show that in the first approach we can obtain the reduced Hamiltonian that
describes the dynamics of remaining degrees of freedom (that are not gauge invariant) and in the second approach the physical 
Hamiltonian that describes the evolution of observables. As pointed out in \cite{Giesel2010} the two Hamiltonians coincide.  

As we have seen in the previous section, the constraints are of the form 
\begin{equation}
\bar{\mathcal{H}}_I = \bar{\mathcal{H}}_I^g + \bar{\mathcal{H}}_I^m\, ,
\end{equation}
where $I\in\{ 0,1\}$, $\bar{\mathcal{H}}_I^g$ is purely gravitational part and
$\bar{\mathcal{H}}_I^m$ is the matter contribution.

Let us start with the case when the kinetic term is absent. 
From the form of constraints (\ref{h0_without_kinetic_term}) and
(\ref{h1_without_kinetic_term}) we see that in
order to obtain the reduced Hamiltonian that describes the dynamics in the reduced phase space, we can choose the following two gauge conditions (that imply two algebraic equations for ${\bar N}$ and  ${\bar N}^1$)
\begin{align}
G_0 &= f-\tau (x,t)\approx 0\, ,\nonumber\\
G_1 &= E_2-\sigma (x)\approx 0\, ,\label{gauge_conditions}
\end{align}
where $\tau$ and $\sigma$ are arbitrary functions, such that $\sigma'(x)\ne 0$,
since, as we have requested earlier $E_2'\ne 0$. This is a good choice for gauge conditions since 
the matrix $M_{IJ}(x,y) := \{ \bar{\mathcal{H}}_I(x),G_J(y)\}$ has a non vanishing determinant in phase space.  Gauge fixing conditions should be invariant under time evolution
\begin{equation}
\frac{\d}{\d t}G_I\approx\frac{\partial}{\partial t}G_I+
\sum_J \int dy\,\bar{N}^J(y)\{ G_I, \bar{\mathcal{H}}_J(y)\}=0\, ,
\end{equation}
so that we obtain
\begin{equation}
{\bar N} = -\frac{\dot{\tau}}{M_{00}}\, , \ \ \ \ {\bar N}^1=0\, ,
\end{equation}
where $M_{IJ}(x)\delta (x-y):=M_{IJ}(x,y)$.

Following the ideas of \cite{Giesel2010}, we see that instead of 
$(\bar{\mathcal{H}}_0,\bar{\mathcal{H}}_1)$ we can 
introduce the set of two locally equivalent constraints $(\tilde{C}_0,\tilde{C}_1)$, lineal in $P_f$ and $K_2$ (since the original constraints contain $K_1'$, there is no local expression for $K_1$),
\begin{align}
 \tilde{C}_0 &:= P_f+h_0(E_1,K_1,E_2,f)\approx 0\, ,
 \label{equivalent_constraints1}\\
 \tilde{C}_1 &:= K_2 + h_1(E_1,K_1,E_2,f)\approx0\, .
 \label{equivalent_constraints2}
\end{align}
As shown in \cite{Thiemann2007} these new constraints are mutually Poisson commuting, $\{\tilde{C}_I,\tilde{C}_J\} =0$.

The Hamiltonian constraint $\bar{\mathcal{H}}_0\approx 0$ is a quadratic equation for
$P_f$, from  (\ref{h0_without_kinetic_term}) it follows that
$\bar{\mathcal{H}}_0 = a P_f^2+bP_f+c$, where $a$, $b$ and $c$ are phase space dependent functions on $(E_1, K_1, E_2, f)$. Then, $h_0$ is one of the solutions of the equation $ah_0^2-bh_0+c=0$.
When we substitute this solution in the diffeomorphism constraint $\mathcal{H}_1\approx 0$ we can find $h_1$. The explicit expressions are given by
\begin{align}
h_0 &= -\frac{1}{E_2'}\biggl(\Sigma_1+\sqrt{(\Sigma_1)^2-\Sigma_2}\biggr)\, 
\label{h_0} ,\\
h_1 &= \frac{1}{E_2'}(f'h_0-K_1'E_1)\, ,
\end{align}
where we have chosen the positive sign in front of the square root, in the solution for $h_0$, and introduced the following notation
\begin{equation}
\Sigma_1 = 4Wf'K_1E_1\, , \ \ \ \ 
\Sigma_2 = 4WE_2'\, [WE_2'f'^2+(E_1)^2\Omega^2\mathcal{H}_0^g]\, .
\end{equation}
Note that $(\tilde{C}_0,\tilde{C}_1)$ define the subset of the full constraint surface, there are two components of the constraint surface corresponding to two possible signs in the solution for $P_f$, each one of them invariant under the gauge transformation. We restrict to the one given by (\ref{h_0}), that is well defined for $E_2'\ne 0$ and $(\Sigma_1)^2-\Sigma_2\ge 0$.


Now, locally and weakly, the relation between the two sets of constraints is given by
\begin{equation}
\bar{\mathcal{H}}_0\approx -M_{00}\tilde{C}_0\, ,\ \ \ \ \  \bar{\mathcal{H}}_1\approx -\sum_I M_{1I}\tilde{C}_I\, ,
\end{equation}
since $M_{00}=-2aP_f-b$, $M_{10}=-f'$ and $M_{11}=E_2'$.
From (\ref{gauge_conditions}) and (\ref{equivalent_constraints1}, \ref{equivalent_constraints2}) we see that $(f,P_f; E_2,K_2)$ can 
be chosen as gauge degrees of freedom, and then the reduced phase space is parameterized by gravitational
degrees of freedom $(E_1,K_1)$. The evolution of an arbitrary function $F(E_1,K_1)$ is defined by
a reduced Hamiltonian, $H_{\rm{red}}$, that produces the same equations of motion as the 
original total Hamiltonian, after fixing the gauge conditions. Then, we obtain 
\begin{equation}
\{  H_{\rm{red}},F\} =  -\int \d x\, {\bar N}M_{00}\{ h_0,F\}\vert_{{\mathcal H}_I=0\, , G_I=0\, ,{\bar N}=-\frac{\dot{\tau}}{M_{00}}}
= \int \d x\, {\dot{\tau}} \{ h_0,F\}\vert_{G_I=0}
\end{equation}
so that
\begin{equation}
 H_{\rm{red}}(E_1,K_1;t) = \int \d x\, {\dot{\tau}}(t)\, h_0 (E_1,K_1;\tau^I)\, ,
 \label{reduced_hamiltonian}
\end{equation}
where $\tau^I=(\tau ,\sigma )$.
The reduced Hamiltonian is explicitly time dependent. 

We can follow the same procedure in the case with the kinetic term, given by (\ref{Total_Hamiltonian_with_kinetic_term}),
where as gauge degrees of freedom we could choose $(E_2,U_2;f,P_f)$ and obtain qualitatively the same results as
in the previous case. 

We have seen that we can parameterize the phase space with canonical coordinates $(E_1,K_1)$, but they are not gauge invariant, since their Poisson brackets with the constraints $\mathcal{H}_I$ do not vanish.
In the gauge invariant approach one constructs the corresponding Dirac observables and analyze their evolution. 
As shown in \cite{Giesel2010}, in the case of coordinate gauge fixing conditions (as in (\ref{gauge_conditions})) the reduced Hamiltonian obtained in the first approach is the same as the physical
Hamiltonian that describes the evolution of observables, associated with arbitrary functions  $F(E_1,K_1)$. These observables are given by
\begin{equation}
O_F(\tau )=[\exp{(X_{\beta})}\cdot F]_{\beta^I=T^I-\tau^I}\,
\label{observables},
\end{equation}
where $T^I=(f,E_2)$, and $X_{\beta}$ is the Hamiltonian vector field of the function
\begin{equation}
\tilde{C}_{\beta}:=\int \d x\, \beta^I \tilde{C}_I\, ,
\end{equation}
where $\beta^I$ do not depend of phase space variables and 
$X_{\beta}\cdot F:=\int \d x\, \beta^I\{ \tilde{C}_I,F\}$. In (\ref{observables})
one first calculates $\exp{(X_{\beta})}\cdot F$ for $\beta$ phase space independent function and only afterwards impose the relation $\beta^I=T^I-\tau^I$. $O_F(\tau )$ is a relational observable, it represents the value of $F$ for $\beta^I=T^I-\tau^I$.  

As shown in \cite{Thiemann2007} (see also references therein) $O_F(\tau )$ are weak Dirac observables with respect to the $\bar{\mathcal{H}}_I$
\begin{equation}
\{ O_F(\tau ),{\bar{\mathcal{H}}}_I\}\approx 0\, .
\end{equation}
There is a Poisson homomorphism, from the commutative algebra of phase space functions to the commutative algebra of relational observables, $F\mapsto O_F(\tau )$, with respect to the Dirac bracket $\{\cdot,\cdot\}_D$ defined by the second class constraints $(\mathcal{H}_I,G_J)$ \cite{Thiemann2007}. Specifically,
\begin{equation}
O_F(\tau )+O_{F'}(\tau )=O_{F+F'}(\tau )\, ,\ \ \ \
O_F(\tau )O_{F'}(\tau )=O_{FF'}(\tau )\, ,
\end{equation}
\begin{equation}
\{ O_F(\tau ),O_{F'}(\tau )\}\approx \{ O_F(\tau ),O_{F'}(\tau )\}_D
\approx O_{\{F,F'\}_D}(\tau )\, .\label{Poisson_hom}
\end{equation}

If we consider a one parameter family of flows $t\mapsto\tau^I(t)$ and define $O_F(t):=O_F(\tau (t))$, then, from (\ref{observables}), we obtain
\begin{equation}
\frac{\d}{\d t}O_F(t)=-\int \d y\, {\dot\tau}^I \sum_{n=0}^{\infty}\, \frac{1}{n!}\int \d x_1
\cdots \int \d x_n\, \beta^{J_1}\cdots\beta^{J_n}X_IX_{J_1}\cdots X_{J_n}\cdot F\, ,
\end{equation}
where $X_J\cdot F:=\{ \tilde{C}_J,F \}$, and the Hamiltonian vector fields $X_J$ commute, as a consequence of $\{\tilde{C}_I, \tilde{C}_J \}=0$.

On the other hand we have
\begin{align}
\{ O_{h_I}(t),O_F(t)\} &= O_{\{ h_I,F \}_D}(t)=O_{\{ h_I,F \}}(t)
= O_{X_I\cdot F}(t) \nonumber\\
&=
\sum_{n=0}^{\infty}\, \frac{1}{n!}\int \d x_1
\cdots \int \d x_n\, \beta^{J_1}\cdots\beta^{J_n}X_IX_{J_1}\cdots X_{J_n}\cdot F\, ,
\end{align}
where $O_{h_I}$ are relational observables associated to $h_I$, given in (\ref{equivalent_constraints2}), and we used the Poisson homomorphism property
(\ref{Poisson_hom}) and since $F=F(E_1,K_1)$ we have $\{ h_I,F \}_D=\{ h_I,F \}$.

From the previous two results follows that we can define the physical Hamiltonian, $H_{\rm{phys}}$, that generates the evolution of observables, by
\begin{equation}
\frac{\d}{\d t}O_F(t)=\{ O_F(t),H_{\rm{phys}}(t)\}\, ,
\end{equation}
where
\begin{equation}
H_{\rm{phys}}(t) := \int \d x\, \dot{\tau}(t)\, O_{h_0}(t)=
\int \d x\,\dot{\tau}(t)\, h_0 (O_{E_1}(t),O_{K_1}(t);\tau^I(t))\, ,
\end{equation}
since $\tau^1=\sigma (x)$. The physical Hamiltonian is
of the same form as the reduced Hamiltonian (\ref{reduced_hamiltonian}), when we identify $F\leftrightarrow O_F(0)$.


\section{Conclusion}
\label{Sec5}

Let us first summarize our results. For a wide class of two dimensional dilatonic models, we considered a new 
set of variables. These are motivated by a reformulation of spherical symmetric models in 4D that make it
suitable for being treated by loop quantization methods. We have extended these `polar type variables' to our 
generic case and defined a set of canonical transformation that render the constraint algebra a true Lie 
algebra. In particular, the Hamiltonian constraints, as defined by different choices of lapse functions, 
commute amongst themselves. We have shown that the resulting Hamiltonian constraint coincides with the one 
that had already been obtained for the metric variables. 
Thus, we recover the clean geometric interpretation that such an 
object has, and that was not obvious to recognize from the polar type variables we started with. 
The new element in this manuscript is that we have extended this result to dilatonic systems with matter,
while previous results had only included the matter-less case.
Finally, we constructed a true Hamiltonian function that plays the role of time evolution generator in the 
reduced phase space picture. Again, this object might be helpful when considering a reduced phase space 
quantization of the models.

Two dimensional models are not only helpful to explore conceptual and technical issues of higher dimensional 
models. Instead, they represent an object of study by themselves. Conformal field theories and their relation 
with holography and string theory have proven to have a very rich structure that justifies being studied in
full detail. It is our belief that having a reformulation of a generic class of such two dimensional models is 
useful. As in any such reformulations, the new description might provide new vistas into the issues at hand 
that might not have been apparent before. Such might be the case of the new variables we have here put forward. 
An obvious first application of this formalism it to attempt to apply loop quantization methods to these 
models,  as has already been done for spherically symmetric gravity and the CGHS model. 
Whether the formulation we have 
here presented might be helpful for unraveling some new structure of these models is an open question that, we believe, deserves further attention.

\begin{acknowledgments}
We would like to thank J.D. Reyes and D. Grumiller for valuable comments. S.R. would like to acknowledge the grant from the 
Programa de Becas Posdoctorales, Centro de Ciencias Matematicas, Campus Morelia, UNAM and DGAPA. S.R. and A.K. would like to thank the support of 
the PROMEP postdoctoral fellowship (through UAM-I), the grant from Sistema Nacional de Investigadores of the 
CONACyT and the partial support of CONACyT grant number 237351: Implicaciones F\'{i}sicas de la Estructura del 
Espaciotiempo. This work was in part supported by DGAPA-UNAM IN103610 grant, by CONACyT 0177840 
and 0232902 grants, by the PASPA-DGAPA program, by NSF PHY-1403943 and PHY-1205388 grants, by CIC, UMSNH grant 
and by the Eberly Research Funds of Penn State.
\end{acknowledgments}

\appendix

\section{The Algebra of Hamiltonian constraint in the vacuum case vs. in the presence of a massless Klein-Gordon field
\label{sub:tot-deriv+matter-append}}

Consider a smeared Hamiltonian constraint of the form $H_{0}(N)=\int \d x\, N(\mathcal{H}_{0}^{g}+\mathcal{H}_{0}^{m})$, 
where $\mathcal{H}_{0}^{g}$ is the pure gravitational-dilaton Hamiltonian constraint and $\mathcal{H}_{0}^{m}$ is the scalar field contribution. 
Computing the Poisson bracket $\{H_{0}(N),H_{0}(M)\}$ yields
\begin{align}
\{H_{0}(N),H_{0}(M)\}= & \left\{ \int \d x\, N(x)\left(\mathcal{H}_{0}^{g}(x)+\mathcal{H}_{0}^{m}(x)\right),\int \d y\, M(y)\left(\mathcal{H}_{0}^{g}(y)+
\mathcal{H}_{0}^{m}(y)\right)\right\} \nonumber \\
= & \int \d x\int \d yN(x)M(y)\left\{ \mathcal{H}_{0}^{g}(x),\mathcal{H}_{0}^{g}(y)\right\} +
\int \d x\int \d yN(x)M(y)\left\{ \mathcal{H}_{0}^{m}(x),\mathcal{H}_{0}^{m}(y)\right\} \nonumber \\
 & +\int \d x\int \d y\left[N(x)M(y)-N(y)M(x)\right]\left\{ \mathcal{H}_{0}^{g}(x),\mathcal{H}_{0}^{m}(y)\right\} .
\label{eq:addit-term}
\end{align}

The original smeared Hamiltonian constraint (\ref{eq:H-noKin-BojoVar}) satisfies the hyper-surface deformation algebra
\begin{equation}
\{ H_{0}(N),H_{0}(M)\}= H_{1}\left( \frac{1}{E_1^2\Omega^2(E_2)}(NM'-MN')\right)\, ,\label{original_hamiltonian_algebra}
\end{equation}
where $H_{1}(N^1)$ is the smeared diffeomorphism constraint. It turns out the gravitational part $H_{0}^{g}(N)$ and the matter part $H_{0}^{m}(N)$ by themselves
satisfy the same Poisson bracket relation (\ref{original_hamiltonian_algebra}) and the last term in (\ref{eq:addit-term}) vanishes.

The rescaling of the lapse and shift functions performed in subsection \ref{Sec3.1}, corresponds to the redefining of the Hamiltonian constraint. The new one is given by
\begin{equation}
\bar{\mathcal{H}}_0:= \frac{E_2'}{E_1\Omega(E_2)}\, \mathcal{H}_0 - \frac{2K_1}{E_1\Omega^2(E_2)}\, \mathcal{H}_1\, .\label{new_lineal_comb}
\end{equation}
We have shown that the gravitational part of this constraint is a spatial derivative of a phase-space function $\mathcal{C}$, see eq.~(\ref{h0_without_kinetic_term}), 
$\bar{\mathcal{H}}_0^g(x)=\partial_x \mathcal{C}(x)$. From (\ref{C}) it follows that $\{ \mathcal{C}(x),\mathcal{C}(y)\} = Q(x,y)\delta (x-y)$, where $Q(x,y)=-Q(y,x)$ (there are no derivatives of the Dirac $\delta$ function). Then, 
the smeared constraint $\bar{H}_{0}^{g}(N)=\int \d x\,N(x)\bar{\mathcal{H}}^g_0(x)$ satisfies
\begin{align}
\{\bar{H}_{0}^{g}(N),\bar{H}_{0}^{g}(M)\}= & \int \d x\int \d y\,N(x)M(y)\left\{ \partial_{x}\mathcal{C}(x),\partial_{y}\mathcal{C}(y)\right\} \nonumber\\
= & \, \int \d x\int \d y N(x)M(y)\partial_{x}\partial_{y}\left[Q(x,y)\delta(x-y)\right]=0\, ,
\end{align}
for arbitrary smearing functions $N(x)$ and $M(y)$, since effectively, as a distribution, $Q(x,y)\delta (x-y)=0$ for every $x$ and $y$,
due to the antisymmetry of $Q(x,y)$. Thus, the Poisson bracket of a smeared total derivative constraint with itself vanishes. 

On the other hand, it is easy to see that the matter part of the new constraint does not become Abelian. Nevertheless, the whole smeared Hamiltonian constraint is Abelian
\begin{equation}
\{ \bar{H}_{0}(N),\bar{H}_{0}(M)\} = 0\, . 
\end{equation}
This can be checked directly by a straightforward calculation, which yields
\begin{equation}
\int \d x\int \d y \bigg( N(x)M(y)\left\{ \bar{\mathcal{H}}_{0}^{m}(x),\bar{\mathcal{H}}_{0}^{m}(y)\right\}
+\left[N(x)M(y)-N(y)M(x)\right]\left\{ \bar{\mathcal{H}}_{0}^{g}(x),\bar{\mathcal{H}}_{0}^{m}(y)\right\}\, \bigg) . \label{eq:T1}\nonumber
\end{equation}
It turns out that in our model, none of these two terms vanishes on its own but they cancel each other and the algebra remains the same 
as in the vacuum case. 

We can trace back the origin of this behaviour by starting from the form of the new Hamiltonian constraint (\ref{new_lineal_comb}), that is a
phase-space dependent linear combination of the original constraints, and analyzing the terms in the Poisson bracket $\{ \bar{H}_{0}(N),\bar{H}_{0}(M)\}$,
taking into account that the vacuum part of this expression vanishes. Then, a long but straightforward analysis shows that all terms cancel
each other, and the algebra of the smeared new Hamiltonian constraint is Abelian.

\section{List of some well-known models that can be written using the generic 2D action}\label{app:models-table}

Here we list some submodels of the generic action (\ref{eq:most-generic-action})\footnote{For a more comprehensive list see  \cite{lineland}. }. Note that in symmetry reduced cases, $\Phi$ is not really the dilaton field but is the factor appearing in front of the $(D-2)$-sphere metric element in the spherically symmetric ansatz, where $D$ is the dimension of the spacetime. 

In what follows, (non-)conf. mean the case for which a conformal transformation has (not) been performed, SS stands for
spherically symmetric and $\lambda$ is the cosmological constant.
\vspace{15pt}

\begin{tabular}{|c|c|c|}
\hline 
Model\vphantom{$\left(\frac{V\left(\left(\nabla\Phi\right)^{2},\Phi\right)}{2}\right)$} & $Y(\Phi)$ & $V\left(\left(\nabla\Phi\right)^{2},\Phi\right)$\tabularnewline
\hline 
\hline 
CGHS (conf.) \citep{C.G.Callan1992, us-new-var,Rastgoo2013} \vphantom{$V\left(\left(\nabla\Phi\right)^{2},\Phi\right)$} & $\frac{1}{8}\Phi^{2}$ & $4\lambda^{2}$\tabularnewline
\hline 
CGHS (non-conf.) \citep{C.G.Callan1992, us-new-var,Rastgoo2013} \vphantom{$V\left(\left(\nabla\Phi\right)^{2},\Phi\right)$} & $\frac{1}{8}\Phi^{2}$ & $\frac{1}{2}\left(\nabla\Phi\right)^{2}+\frac{1}{2}\lambda^{2}\Phi^{2}$\tabularnewline
\hline 
3+1 SS without $\lambda$ (conf.) \citep{us-new-var,Rastgoo2013} \vphantom{$V\left(\left(\nabla\Phi\right)^{2},\Phi\right)$} & $\frac{1}{4}\Phi^{2}$ & $\frac{1}{2\Phi}$\tabularnewline
\hline 
3+1 SS without $\lambda$ (non-conf.) \citep{us-new-var,Rastgoo2013} \vphantom{$V\left(\left(\nabla\Phi\right)^{2},\Phi\right)$} & $\frac{1}{4}\Phi^{2}$ & $\frac{1}{2}\left(\nabla\Phi\right)^{2}+\frac{1}{2}$\tabularnewline
\hline 
$D$-dim. SS  with $\lambda$ (non-conf.) \citep{Grumiller2002} \vphantom{$\left(\frac{\left(\frac{V\left(\left(\nabla\Phi\right)^{2},\Phi\right)}{2}\right)}{a}\right)$} & $\frac{1}{4}\Phi^{2}$ & $\frac{-4(D-3)}{\Phi^{2}(D-2)}\left(\nabla\Phi\right)^{2}+\frac{\lambda^{2}}{4}(D-2)(D-3)\left(\frac{\Phi^{2}}{4}\right)^{\frac{D-4}{D-2}}$\tabularnewline
\hline 
Liouville gravity \citep{Liouville,Grumiller2002} \vphantom{$V\left(\left(\nabla\Phi\right)^{2},\Phi\right)$} & $(b+b^{-1})\Phi$ & $\left(\nabla\Phi\right)^{2}+4\pi\mu e^{2b\Phi}$\tabularnewline
\hline 
Jackiw, Teitelboim 
model \citep{Teitel83,Jackiw85, Jackiw,Grumiller2002} \vphantom{$V\left(\left(\nabla\Phi\right)^{2},\Phi\right)$} & $\Phi$ & $\frac{1}{2}\left(\nabla\Phi\right)^{2}+\Lambda$\tabularnewline
\hline 
\end{tabular}
\vspace{15pt}


\section{Form of $\mathcal{H}_{0}^{g}$: Comparison with metric formalism}
\label{app:comparison}

As we have seen in (\ref{h0_without_kinetic_term}) the matter-less part of the Hamiltonian constraint,
$\bar{\mathcal{H}}_{0}^{g}$, in the case when the kinetic term is absent, can be written as a spatial derivative 
\begin{eqnarray}
\label{total-dev}
\bar{\mathcal{H}}_{0}^{g}:=\mathcal C'=\partial_x\bigg(\frac{E_2'^2}{4E^2_1\Omega^2(E_2)}-\frac{K_1^2}{\Omega^2(E_2)}
-\int V(E_2)\textrm{d}E_2\bigg).
\end{eqnarray}
This is the analogue of the result of \cite{Louis-Martinez1994}, obtained in metric variables, and we shall show that the
two expressions are equivalent.
In \cite{Louis-Martinez1994}, configuration variables are $\rho$ and $\phi$, where $\rho$ is defined by the following
parameterization of the metric
\begin{equation}
\d s^2= e^{2\rho}[-N^2 \d t^2 + (\d x + N^1 \d t)^2]\, ,
\end{equation}
and $\phi =Y(\Phi )$.

The relations between these variables and $(E_1,E_2)$ are given by 
\begin{eqnarray}
E_1&=&\frac{e^{\rho}}{\bar\Omega(\phi)}\\
E_2&=&4\phi,
\end{eqnarray}
where $\bar\Omega(\phi)=\Omega(E_2(\phi))$.

Therefore, we have the following generating function
\begin{equation}
G(Q,p)=\frac{e^{\rho}}{\bar\Omega(\phi)}K_1+4\phi K_2,
\end{equation}
so the momenta related to $\rho$ and $\phi$ are
\begin{eqnarray}
\Pi_\rho&=&\frac{e^{\rho}}{\bar\Omega(\phi)}K_1=E_1K_1\\
\Pi_\phi&=&-\frac{1}{\bar\Omega(\phi)}\frac{\textrm{d}\bar\Omega ({\phi})}{\textrm{d}\phi}\, E_1K_1+4K_2.
\end{eqnarray}
Then, $K_1$ in terms of these new variables is
\begin{equation}
K_1=e^{-\rho}\Pi_\rho\bar\Omega(\phi).
\end{equation}
By substituting these new variables in (\ref{total-dev}) we get
\begin{equation}
\mathcal C= 4\bigg[e^{-2\rho}\bigg(\phi'^2-\frac{1}{4}\Pi_\rho^2\bigg)
-\int \bar V(\phi)\textrm{d}\phi\bigg],
\end{equation}
where $\bar V(\phi)=V(E_2(\phi))$.

In \cite{Louis-Martinez1994},  the Hamiltonian constraint is rewritten as a spatial derivative of
\begin{equation}
\mathcal C_1=e^{-2\rho}\bigg(\frac{1}{4}\Pi_\rho^2-\phi'^2\bigg)-j(\phi)
\end{equation}
where $\textrm{d}j(\phi)/\textrm{d}\phi=-\bar V(\phi)$.
Therefore, we see that $\mathcal C=-4\mathcal C_1$, so ours is essentially the same expression, written in a different  set of canonical variables.


\bibliography{Bib1+1}

\begin{thebibliography}{28}%
\makeatletter
\providecommand \@ifxundefined [1]{%
 \@ifx{#1\undefined}
}%
\providecommand \@ifnum [1]{%
 \ifnum #1\expandafter \@firstoftwo
 \else \expandafter \@secondoftwo
 \fi
}%
\providecommand \@ifx [1]{%
 \ifx #1\expandafter \@firstoftwo
 \else \expandafter \@secondoftwo
 \fi
}%
\providecommand \natexlab [1]{#1}%
\providecommand \enquote  [1]{``#1''}%
\providecommand \bibnamefont  [1]{#1}%
\providecommand \bibfnamefont [1]{#1}%
\providecommand \citenamefont [1]{#1}%
\providecommand \href@noop [0]{\@secondoftwo}%
\providecommand \href [0]{\begingroup \@sanitize@url \@href}%
\providecommand \@href[1]{\@@startlink{#1}\@@href}%
\providecommand \@@href[1]{\endgroup#1\@@endlink}%
\providecommand \@sanitize@url [0]{\catcode `\\12\catcode `\$12\catcode
  `\&12\catcode `\#12\catcode `\^12\catcode `\_12\catcode `\%12\relax}%
\providecommand \@@startlink[1]{}%
\providecommand \@@endlink[0]{}%
\providecommand \url  [0]{\begingroup\@sanitize@url \@url }%
\providecommand \@url [1]{\endgroup\@href {#1}{\urlprefix }}%
\providecommand \urlprefix  [0]{URL }%
\providecommand \Eprint [0]{\href }%
\providecommand \doibase [0]{http://dx.doi.org/}%
\providecommand \selectlanguage [0]{\@gobble}%
\providecommand \bibinfo  [0]{\@secondoftwo}%
\providecommand \bibfield  [0]{\@secondoftwo}%
\providecommand \translation [1]{[#1]}%
\providecommand \BibitemOpen [0]{}%
\providecommand \bibitemStop [0]{}%
\providecommand \bibitemNoStop [0]{.\EOS\space}%
\providecommand \EOS [0]{\spacefactor3000\relax}%
\providecommand \BibitemShut  [1]{\csname bibitem#1\endcsname}%
\let\auto@bib@innerbib\@empty
\bibitem [{\citenamefont {Callan}\ \emph {et~al.}(1992)\citenamefont {Callan},
  \citenamefont {Giddings}, \citenamefont {Harvey},\ and\ \citenamefont
  {Strominger}}]{C.G.Callan1992}%
  \BibitemOpen
  \bibfield  {author} {\bibinfo {author} {\bibfnamefont {C.~G.}\ \bibnamefont
  {Callan}}, \bibinfo {author} {\bibfnamefont {S.~B.}\ \bibnamefont
  {Giddings}}, \bibinfo {author} {\bibfnamefont {J.~A.}\ \bibnamefont
  {Harvey}}, \ and\ \bibinfo {author} {\bibfnamefont {A.}~\bibnamefont
  {Strominger}},\ }\bibfield  {title} {\enquote {\bibinfo {title} {Evanescent
  black holes},}\ }\href@noop {} {\bibfield  {journal} {\bibinfo  {journal}
  {Phys. Rev.}\ }\textbf {\bibinfo {volume} {D45}},\ \bibinfo {pages} {R1005}
  (\bibinfo {year} {1992})},\ \Eprint
  {http://arxiv.org/abs/arXiv:hep-th/9111056v1} {arXiv:hep-th/9111056v1}
  \BibitemShut {NoStop}%
\bibitem [{\citenamefont {Kuchar}(1994)}]{Kuchar1994}%
  \BibitemOpen
  \bibfield  {author} {\bibinfo {author} {\bibfnamefont {K.~V.}\ \bibnamefont
  {Kuchar}},\ }\bibfield  {title} {\enquote {\bibinfo {title} {Geometrodynamics
  of {S}chwarzschild black holes},}\ }\href {\doibase 10.1103/PhysRevD.50.3961}
  {\bibfield  {journal} {\bibinfo  {journal} {Phys.Rev.}\ }\textbf {\bibinfo
  {volume} {D50}},\ \bibinfo {pages} {3961} (\bibinfo {year} {1994})},\ \Eprint
  {http://arxiv.org/abs/gr-qc/9403003} {arXiv:gr-qc/9403003} \BibitemShut
  {NoStop}%
\bibitem [{\citenamefont {Grumiller}\ \emph {et~al.}(2002)\citenamefont
  {Grumiller}, \citenamefont {Kummer},\ and\ \citenamefont
  {Vassilevich}}]{Grumiller2002}%
  \BibitemOpen
  \bibfield  {author} {\bibinfo {author} {\bibfnamefont {D.}~\bibnamefont
  {Grumiller}}, \bibinfo {author} {\bibfnamefont {W.}~\bibnamefont {Kummer}}, \
  and\ \bibinfo {author} {\bibfnamefont {D.~V.}\ \bibnamefont {Vassilevich}},\
  }\bibfield  {title} {\enquote {\bibinfo {title} {Dilaton gravity in
  two-dimensions},}\ }\href@noop {} {\bibfield  {journal} {\bibinfo  {journal}
  {Phys. Rept.}\ }\textbf {\bibinfo {volume} {369}},\ \bibinfo {pages} {327}
  (\bibinfo {year} {2002})},\ \Eprint
  {http://arxiv.org/abs/arXiv:hep-th/0204253} {arXiv:hep-th/0204253}
  \BibitemShut {NoStop}%
\bibitem [{\citenamefont {Kloesch}\ and\ \citenamefont
  {Strobl}(1996)}]{Kloesch1996}%
  \BibitemOpen
  \bibfield  {author} {\bibinfo {author} {\bibfnamefont {T.}~\bibnamefont
  {Kloesch}}\ and\ \bibinfo {author} {\bibfnamefont {T.}~\bibnamefont
  {Strobl}},\ }\bibfield  {title} {\enquote {\bibinfo {title} {Classical and
  quantum gravity in 1+1 dimensions, part {I}: A unifying approach},}\
  }\href@noop {} {\bibfield  {journal} {\bibinfo  {journal} {Class. Quantum
  Grav.}\ }\textbf {\bibinfo {volume} {13}},\ \bibinfo {pages} {965} (\bibinfo
  {year} {1996})},\ \Eprint {http://arxiv.org/abs/arXiv:gr-qc/9508020}
  {arXiv:gr-qc/9508020} \BibitemShut {NoStop}%
\bibitem [{\citenamefont {Kummer}\ and\ \citenamefont
  {Lau}(1997)}]{Kummer_Lau1997}%
  \BibitemOpen
  \bibfield  {author} {\bibinfo {author} {\bibfnamefont {W.}~\bibnamefont
  {Kummer}}\ and\ \bibinfo {author} {\bibfnamefont {S.~R.}\ \bibnamefont
  {Lau}},\ }\bibfield  {title} {\enquote {\bibinfo {title} {Boundary conditions
  and quasilocal energy in the canonical formulation of all 1+1 models of
  gravity},}\ }\href@noop {} {\bibfield  {journal} {\bibinfo  {journal} {Ann.
  Phys.}\ }\textbf {\bibinfo {volume} {258}},\ \bibinfo {pages} {37} (\bibinfo
  {year} {1997})},\ \Eprint {http://arxiv.org/abs/arXiv:gr-qc/9612021}
  {arXiv:gr-qc/9612021} \BibitemShut {NoStop}%
\bibitem [{\citenamefont {Varadarajan}(1995)}]{Varadaraj95}%
  \BibitemOpen
  \bibfield  {author} {\bibinfo {author} {\bibfnamefont {M.}~\bibnamefont
  {Varadarajan}},\ }\bibfield  {title} {\enquote {\bibinfo {title} {Classical
  and quantum geometrodynamics of 2-d vacuum dilatonic black holes},}\
  }\href@noop {} {\bibfield  {journal} {\bibinfo  {journal} {Phys. Rev.}\
  }\textbf {\bibinfo {volume} {D 52}},\ \bibinfo {pages} {7080--7088} (\bibinfo
  {year} {1995})},\ \Eprint {http://arxiv.org/abs/arXiv:gr-qc/9508039}
  {arXiv:gr-qc/9508039} \BibitemShut {NoStop}%
\bibitem [{\citenamefont {Louis-Martinez}\ \emph {et~al.}(1994)\citenamefont
  {Louis-Martinez}, \citenamefont {Gegenberg},\ and\ \citenamefont
  {Kunstatter}}]{Louis-Martinez1994}%
  \BibitemOpen
  \bibfield  {author} {\bibinfo {author} {\bibfnamefont {D.}~\bibnamefont
  {Louis-Martinez}}, \bibinfo {author} {\bibfnamefont {J.}~\bibnamefont
  {Gegenberg}}, \ and\ \bibinfo {author} {\bibfnamefont {G.}~\bibnamefont
  {Kunstatter}},\ }\bibfield  {title} {\enquote {\bibinfo {title} {Exact
  {D}irac quantization of all 2d dilaton gravity theory},}\ }\href@noop {}
  {\bibfield  {journal} {\bibinfo  {journal} {Phys. Lett.}\ }\textbf {\bibinfo
  {volume} {B 321}},\ \bibinfo {pages} {193} (\bibinfo {year} {1994})},\
  \Eprint {http://arxiv.org/abs/arXiv:gr-qc/9309018} {arXiv:gr-qc/9309018}
  \BibitemShut {NoStop}%
\bibitem [{\citenamefont {Louis-Martinez}(1997)}]{Louis-Martinez1997}%
  \BibitemOpen
  \bibfield  {author} {\bibinfo {author} {\bibfnamefont {D.}~\bibnamefont
  {Louis-Martinez}},\ }\bibfield  {title} {\enquote {\bibinfo {title} {Dirac
  quantization of two-dimensional dilation gravity minimally coupled to {N}
  massless scalar fields},}\ }\href@noop {} {\bibfield  {journal} {\bibinfo
  {journal} {Phys. Rev.}\ }\textbf {\bibinfo {volume} {D 55}},\ \bibinfo
  {pages} {7982} (\bibinfo {year} {1997})},\ \Eprint
  {http://arxiv.org/abs/arXiv:hep-th/9611031} {arXiv:hep-th/9611031}
  \BibitemShut {NoStop}%
\bibitem [{\citenamefont {Kuchar}\ \emph {et~al.}(1997)\citenamefont {Kuchar},
  \citenamefont {Romano},\ and\ \citenamefont {Varadarajan}}]{Kuchar1997}%
  \BibitemOpen
  \bibfield  {author} {\bibinfo {author} {\bibfnamefont {K.~V.}\ \bibnamefont
  {Kuchar}}, \bibinfo {author} {\bibfnamefont {J.~D.}\ \bibnamefont {Romano}},
  \ and\ \bibinfo {author} {\bibfnamefont {M.}~\bibnamefont {Varadarajan}},\
  }\bibfield  {title} {\enquote {\bibinfo {title} {Dirac constraint
  quantization of a dilatonic model of gravitational collapse},}\ }\href@noop
  {} {\bibfield  {journal} {\bibinfo  {journal} {Phys. Rev.}\ }\textbf
  {\bibinfo {volume} {D 55}},\ \bibinfo {pages} {795} (\bibinfo {year}
  {1997})},\ \Eprint {http://arxiv.org/abs/arXiv:gr-qc/9608011}
  {arXiv:gr-qc/9608011} \BibitemShut {NoStop}%
\bibitem [{\citenamefont {Ashtekar}\ \emph {et~al.}(2011)\citenamefont
  {Ashtekar}, \citenamefont {Pretorius},\ and\ \citenamefont
  {Ramazanoglu}}]{A.Ashtekar2011}%
  \BibitemOpen
  \bibfield  {author} {\bibinfo {author} {\bibfnamefont {A.}~\bibnamefont
  {Ashtekar}}, \bibinfo {author} {\bibfnamefont {F.}~\bibnamefont {Pretorius}},
  \ and\ \bibinfo {author} {\bibfnamefont {F.~M.}\ \bibnamefont
  {Ramazanoglu}},\ }\bibfield  {title} {\enquote {\bibinfo {title} {Evaporation
  of 2-dimensional black holes},}\ }\href@noop {} {\bibfield  {journal}
  {\bibinfo  {journal} {Phys. Rev}\ }\textbf {\bibinfo {volume} {D83}},\
  \bibinfo {pages} {044040} (\bibinfo {year} {2011})},\ \Eprint
  {http://arxiv.org/abs/arXiv:1012.0077 [gr-qc]} {arXiv:1012.0077 [gr-qc]}
  \BibitemShut {NoStop}%
\bibitem [{\citenamefont {Gambini}\ and\ \citenamefont
  {Pullin}(2014)}]{Gambini_Pullin2014}%
  \BibitemOpen
  \bibfield  {author} {\bibinfo {author} {\bibfnamefont {R.}~\bibnamefont
  {Gambini}}\ and\ \bibinfo {author} {\bibfnamefont {J.}~\bibnamefont
  {Pullin}},\ }\bibfield  {title} {\enquote {\bibinfo {title} {Hawking
  radiation from a spherical loop quantum gravity black hole},}\ }\href@noop {}
  {\bibfield  {journal} {\bibinfo  {journal} {Class. Quant. Grav.}\ }\textbf
  {\bibinfo {volume} {31}},\ \bibinfo {pages} {115003} (\bibinfo {year}
  {2014})},\ \Eprint {http://arxiv.org/abs/arXiv:1312.3595 [gr-qc]}
  {arXiv:1312.3595 [gr-qc]} \BibitemShut {NoStop}%
\bibitem [{\citenamefont {Bojowald}\ and\ \citenamefont
  {Swiderski}(2006)}]{BSvars05}%
  \BibitemOpen
  \bibfield  {author} {\bibinfo {author} {\bibfnamefont {M.}~\bibnamefont
  {Bojowald}}\ and\ \bibinfo {author} {\bibfnamefont {R.}~\bibnamefont
  {Swiderski}},\ }\bibfield  {title} {\enquote {\bibinfo {title} {Spherically
  symmetric quantum geometry: Hamiltonian constraint},}\ }\href@noop {}
  {\bibfield  {journal} {\bibinfo  {journal} {Class. Quantum Grav.}\ }\textbf
  {\bibinfo {volume} {23}},\ \bibinfo {pages} {2129} (\bibinfo {year}
  {2006})},\ \Eprint {http://arxiv.org/abs/arXiv:gr-qc/0511108}
  {arXiv:gr-qc/0511108} \BibitemShut {NoStop}%
\bibitem [{\citenamefont {Gambini}\ \emph {et~al.}(2010)\citenamefont
  {Gambini}, \citenamefont {Pullin},\ and\ \citenamefont
  {Rastgoo}}]{us-new-var}%
  \BibitemOpen
  \bibfield  {author} {\bibinfo {author} {\bibfnamefont {R.}~\bibnamefont
  {Gambini}}, \bibinfo {author} {\bibfnamefont {J.}~\bibnamefont {Pullin}}, \
  and\ \bibinfo {author} {\bibfnamefont {S.}~\bibnamefont {Rastgoo}},\
  }\bibfield  {title} {\enquote {\bibinfo {title} {New variables for 1+1
  dimensional gravity},}\ }\href@noop {} {\bibfield  {journal} {\bibinfo
  {journal} {Class. Quantum Grav.}\ }\textbf {\bibinfo {volume} {27}},\
  \bibinfo {pages} {025002} (\bibinfo {year} {2010})},\ \Eprint
  {http://arxiv.org/abs/arXiv:0909.0459v2 [gr-qc]} {arXiv:0909.0459v2 [gr-qc]}
  \BibitemShut {NoStop}%
\bibitem [{\citenamefont {Gambini}\ and\ \citenamefont
  {Pullin}(2013)}]{Gamb_Pull2013}%
  \BibitemOpen
  \bibfield  {author} {\bibinfo {author} {\bibfnamefont {R.}~\bibnamefont
  {Gambini}}\ and\ \bibinfo {author} {\bibfnamefont {J.}~\bibnamefont
  {Pullin}},\ }\bibfield  {title} {\enquote {\bibinfo {title} {Loop
  quantization of the {S}chwarzschild black hole},}\ }\href@noop {} {\bibfield
  {journal} {\bibinfo  {journal} {Phys. Rev. Lett.}\ }\textbf {\bibinfo
  {volume} {110}},\ \bibinfo {pages} {211301} (\bibinfo {year} {2013})},\
  \Eprint {http://arxiv.org/abs/arXiv:1302.5265 [gr-qc]} {arXiv:1302.5265
  [gr-qc]} \BibitemShut {NoStop}%
\bibitem [{\citenamefont {Rastgoo}(2013)}]{Rastgoo2013}%
  \BibitemOpen
  \bibfield  {author} {\bibinfo {author} {\bibfnamefont {S.}~\bibnamefont
  {Rastgoo}},\ }\bibfield  {title} {\enquote {\bibinfo {title} {A local true
  {H}amiltonian for the {CGHS} model in new variables},}\ }\href@noop {} {\
  (\bibinfo {year} {2013})},\ \Eprint {http://arxiv.org/abs/arXiv:1304.7836
  [gr-qc]} {arXiv:1304.7836 [gr-qc]} \BibitemShut {NoStop}%
\bibitem [{\citenamefont {Rovelli}(1991)}]{Rovelli1991}%
  \BibitemOpen
  \bibfield  {author} {\bibinfo {author} {\bibfnamefont {C.}~\bibnamefont
  {Rovelli}},\ }\bibfield  {title} {\enquote {\bibinfo {title} {What is
  observable in classical and quantum gravity?}}\ }\href@noop {} {\bibfield
  {journal} {\bibinfo  {journal} {Class. Quantum Grav.}\ }\textbf {\bibinfo
  {volume} {8}},\ \bibinfo {pages} {1895} (\bibinfo {year} {1991})}\BibitemShut
  {NoStop}%
\bibitem [{\citenamefont {Giesel}\ \emph {et~al.}(2010)\citenamefont {Giesel},
  \citenamefont {Hofmann}, \citenamefont {Thiemann},\ and\ \citenamefont
  {Winkler}}]{Giesel2010}%
  \BibitemOpen
  \bibfield  {author} {\bibinfo {author} {\bibfnamefont {K.}~\bibnamefont
  {Giesel}}, \bibinfo {author} {\bibfnamefont {S.}~\bibnamefont {Hofmann}},
  \bibinfo {author} {\bibfnamefont {T.}~\bibnamefont {Thiemann}}, \ and\
  \bibinfo {author} {\bibfnamefont {O.}~\bibnamefont {Winkler}},\ }\bibfield
  {title} {\enquote {\bibinfo {title} {Manifestly gauge-invariant general
  relativistic perturbation theory : {I}. {F}oundations},}\ }\href@noop {}
  {\bibfield  {journal} {\bibinfo  {journal} {Class. Quant. Grav.}\ }\textbf
  {\bibinfo {volume} {27}},\ \bibinfo {pages} {055005} (\bibinfo {year}
  {2010})},\ \Eprint {http://arxiv.org/abs/arXiv:0711.0115 [gr-qc]}
  {arXiv:0711.0115 [gr-qc]} \BibitemShut {NoStop}%
\bibitem [{\citenamefont {Giesel}\ and\ \citenamefont
  {Thiemann}(2010)}]{Thiemann2007}%
  \BibitemOpen
  \bibfield  {author} {\bibinfo {author} {\bibfnamefont {K.}~\bibnamefont
  {Giesel}}\ and\ \bibinfo {author} {\bibfnamefont {T.}~\bibnamefont
  {Thiemann}},\ }\bibfield  {title} {\enquote {\bibinfo {title} {Algebraic
  quantum gravity ({AQG}) {IV}. {R}educed phase space quantisation of loop
  quantum gravity},}\ }\href@noop {} {\bibfield  {journal} {\bibinfo  {journal}
  {Class. Quant. Grav.}\ }\textbf {\bibinfo {volume} {27}},\ \bibinfo {pages}
  {175009} (\bibinfo {year} {2010})},\ \Eprint
  {http://arxiv.org/abs/arXiv:0711.0119 [gr-qc]} {arXiv:0711.0119 [gr-qc]}
  \BibitemShut {NoStop}%
\bibitem [{\citenamefont {Grumiller}(2003)}]{Grumiller_PhD}%
  \BibitemOpen
  \bibfield  {author} {\bibinfo {author} {\bibfnamefont {D.}~\bibnamefont
  {Grumiller}},\ }\bibfield  {title} {\enquote {\bibinfo {title} {Quantum
  dilaton gravity in two dimensions with matter},}\ }\href@noop {} {\bibfield
  {journal} {\bibinfo  {journal} {PhD Thesis}\ } (\bibinfo {year} {2003})},\
  \Eprint {http://arxiv.org/abs/arXiv:gr-qc/0105078v4} {arXiv:gr-qc/0105078v4}
  \BibitemShut {NoStop}%
\bibitem [{\citenamefont {Banks}\ and\ \citenamefont
  {O'Loughlin}(1991)}]{Banks1991}%
  \BibitemOpen
  \bibfield  {author} {\bibinfo {author} {\bibfnamefont {T.}~\bibnamefont
  {Banks}}\ and\ \bibinfo {author} {\bibfnamefont {M.}~\bibnamefont
  {O'Loughlin}},\ }\bibfield  {title} {\enquote {\bibinfo {title}
  {Two-dimensional quantum gravity in {M}inkowski space},}\ }\href@noop {}
  {\bibfield  {journal} {\bibinfo  {journal} {Nucl. Phys.}\ }\textbf {\bibinfo
  {volume} {B362}},\ \bibinfo {pages} {649} (\bibinfo {year}
  {1991})}\BibitemShut {NoStop}%
\bibitem [{\citenamefont {Odintsov}\ and\ \citenamefont
  {Shapiro}(1991)}]{Odintsov1991}%
  \BibitemOpen
  \bibfield  {author} {\bibinfo {author} {\bibfnamefont {S.D.}\ \bibnamefont
  {Odintsov}}\ and\ \bibinfo {author} {\bibfnamefont {I.L.}\ \bibnamefont
  {Shapiro}},\ }\bibfield  {title} {\enquote {\bibinfo {title} {One-loop
  renormalization of two-dimensional induced quantum gravity},}\ }\href@noop {}
  {\bibfield  {journal} {\bibinfo  {journal} {Phys. Lett.}\ }\textbf {\bibinfo
  {volume} {B263}},\ \bibinfo {pages} {183} (\bibinfo {year}
  {1991})}\BibitemShut {NoStop}%
\bibitem [{\citenamefont {Gegenberg}\ \emph {et~al.}(1995)\citenamefont
  {Gegenberg}, \citenamefont {Kunstatter},\ and\ \citenamefont
  {Louis-Martinez}}]{Gegenberg1995}%
  \BibitemOpen
  \bibfield  {author} {\bibinfo {author} {\bibfnamefont {J.}~\bibnamefont
  {Gegenberg}}, \bibinfo {author} {\bibfnamefont {G.}~\bibnamefont
  {Kunstatter}}, \ and\ \bibinfo {author} {\bibfnamefont {D.}~\bibnamefont
  {Louis-Martinez}},\ }\bibfield  {title} {\enquote {\bibinfo {title}
  {Observables for two-dimensional black holes},}\ }\href@noop {} {\bibfield
  {journal} {\bibinfo  {journal} {Phys. Rev.}\ }\textbf {\bibinfo {volume} {D
  51}},\ \bibinfo {pages} {1781} (\bibinfo {year} {1995})},\ \Eprint
  {http://arxiv.org/abs/arXiv:gr-qc/9408015} {arXiv:gr-qc/9408015} \BibitemShut
  {NoStop}%
\bibitem [{\citenamefont {Grumiller}\ and\ \citenamefont
  {McNees}(2007)}]{Grumiller_McNees}%
  \BibitemOpen
  \bibfield  {author} {\bibinfo {author} {\bibfnamefont {D.}~\bibnamefont
  {Grumiller}}\ and\ \bibinfo {author} {\bibfnamefont {R.}~\bibnamefont
  {McNees}},\ }\bibfield  {title} {\enquote {\bibinfo {title} {Thermodynamics
  of black holes in two (and higher) dimensions},}\ }\href@noop {} {\bibfield
  {journal} {\bibinfo  {journal} {JHEP}\ }\textbf {\bibinfo {volume} {04}},\
  \bibinfo {pages} {074} (\bibinfo {year} {2007})},\ \Eprint
  {http://arxiv.org/abs/arXiv:hep-th/0703230} {arXiv:hep-th/0703230}
  \BibitemShut {NoStop}%
\bibitem [{\citenamefont {Grumiller}\ and\ \citenamefont
  {Meyer}(2006)}]{lineland}%
  \BibitemOpen
  \bibfield  {author} {\bibinfo {author} {\bibfnamefont {D.}~\bibnamefont
  {Grumiller}}\ and\ \bibinfo {author} {\bibfnamefont {R.}~\bibnamefont
  {Meyer}},\ }\bibfield  {title} {\enquote {\bibinfo {title} {Ramifications of
  lineland},}\ }\href@noop {} {\bibfield  {journal} {\bibinfo  {journal} {Turk.
  J. Phys.}\ }\textbf {\bibinfo {volume} {30}},\ \bibinfo {pages} {349}
  (\bibinfo {year} {2006})},\ \Eprint
  {http://arxiv.org/abs/arXiv:hep-th/0604049} {arXiv:hep-th/0604049}
  \BibitemShut {NoStop}%
\bibitem [{\citenamefont {D'Hoker}\ and\ \citenamefont
  {Jackiw}(1982)}]{Liouville}%
  \BibitemOpen
  \bibfield  {author} {\bibinfo {author} {\bibfnamefont {E.}~\bibnamefont
  {D'Hoker}}\ and\ \bibinfo {author} {\bibfnamefont {R.}~\bibnamefont
  {Jackiw}},\ }\bibfield  {title} {\enquote {\bibinfo {title} {Classical and
  quantal {L}iouville field theory},}\ }\href@noop {} {\bibfield  {journal}
  {\bibinfo  {journal} {Phys. Rev.}\ }\textbf {\bibinfo {volume} {D 26}},\
  \bibinfo {pages} {3517} (\bibinfo {year} {1982})}\BibitemShut {NoStop}%
\bibitem [{\citenamefont {Teitelboim}(1983)}]{Teitel83}%
  \BibitemOpen
  \bibfield  {author} {\bibinfo {author} {\bibfnamefont {C.}~\bibnamefont
  {Teitelboim}},\ }\bibfield  {title} {\enquote {\bibinfo {title} {Supergravity
  and hamiltonian structure in two spacetime dimensions},}\ }\href@noop {}
  {\bibfield  {journal} {\bibinfo  {journal} {Phys. Lett.}\ }\textbf {\bibinfo
  {volume} {B 126}},\ \bibinfo {pages} {46} (\bibinfo {year}
  {1983})}\BibitemShut {NoStop}%
\bibitem [{\citenamefont {Jackiw}(1985)}]{Jackiw85}%
  \BibitemOpen
  \bibfield  {author} {\bibinfo {author} {\bibfnamefont {R.}~\bibnamefont
  {Jackiw}},\ }\bibfield  {title} {\enquote {\bibinfo {title} {Lower
  dimensional gravity},}\ }\href@noop {} {\bibfield  {journal} {\bibinfo
  {journal} {Nucl. Phys.}\ }\textbf {\bibinfo {volume} {B 252}},\ \bibinfo
  {pages} {343} (\bibinfo {year} {1985})}\BibitemShut {NoStop}%
\bibitem [{\citenamefont {Mann}\ \emph {et~al.}(1990)\citenamefont {Mann},
  \citenamefont {Shiekh},\ and\ \citenamefont {Tarasov}}]{Jackiw}%
  \BibitemOpen
  \bibfield  {author} {\bibinfo {author} {\bibfnamefont {R.}~\bibnamefont
  {Mann}}, \bibinfo {author} {\bibfnamefont {A.}~\bibnamefont {Shiekh}}, \ and\
  \bibinfo {author} {\bibfnamefont {L.}~\bibnamefont {Tarasov}},\ }\bibfield
  {title} {\enquote {\bibinfo {title} {Classical and quantum properties of
  two-dimensional black holes},}\ }\href@noop {} {\bibfield  {journal}
  {\bibinfo  {journal} {Nucl. Phys.}\ }\textbf {\bibinfo {volume} {B 341}},\
  \bibinfo {pages} {134} (\bibinfo {year} {1990})}\BibitemShut {NoStop}%
\end{thebibliography}%

\end{document}